\newcommand{\ARTICLE}

\ifdefined\ARTICLE
\documentclass[11pt]{article}
\else
\documentclass[11pt]{svjour3}
\fi

\usepackage{amsmath, amsfonts, amssymb}
\usepackage{graphicx}
\usepackage{subfigure}
\usepackage{hyperref}
\usepackage[small]{caption}
\usepackage{appendix}

\ifdefined\ARTICLE
\usepackage{amsthm}
\usepackage{fullpage}
\fi

\hypersetup{
    colorlinks=false,
    pdfborder={0 0 0},
}

\newcommand{\jImageWidth}{120mm}
\newcommand{\jImageHeight}{70mm}

\ifdefined\ARTICLE
\theoremstyle{plain}
\newtheorem{theorem}{Theorem}[section]
\newtheorem{lemma}[theorem]{Lemma}
\newtheorem{proposition}[theorem]{Proposition}

\newtheorem{definition}[theorem]{Definition}
\newtheorem{remark}[theorem]{Remark}

\fi

\newcommand{\ceiling}[1]{\left\lceil{#1}\right\rceil}
\newcommand{\floor}[1]{\left\lfloor{#1}\right\rfloor}
\newcommand{\nfrac}[2]{\left( \frac{#1}{#2} \right)}
\newcommand{\brac}[1]{\left\{ #1 \right\}}
\newcommand{\bras}[1]{\left[ #1 \right]}
\newcommand{\brap}[1]{\left( #1 \right)}
\newcommand{\fpow}[3]{\left(\frac{#1}{#2}\right)^{#3}}

\newcommand{\Exp}{\mathbb{E}}
\newcommand{\Expp}{\mathbb{E}_{\pi}}

\newcommand{\pig}{\pi_{\gamma}}
\newcommand{\sZ}{\mathbb{Z}_{+}}
\newcommand{\sR}{\mathbb{R}_{+}}

\newcommand{\Deq}{\stackrel{\Delta} = }
\newcommand{\Qg}{\overline{Q}(\gamma)}

\newcommand{\Cg}{\overline{C}(\gamma)}
\newcommand{\Ug}{\overline{U}(\gamma)}

\newcommand{\Qgk}{\overline{Q}(\gamma_{k})}

\newcommand{\Ugk}{\overline{U}(\gamma_{k})}

\newcommand{\Cgk}{\overline{C}(\gamma_{k})}
\newcommand{\XL}{\mathcal{X}_{\lambda}}
\newcommand{\XM}{\mathcal{X}_{\mu}}
\newcommand{\INTMC}{$\mu$-CHOICE}
\newcommand{\INTLC}{$\lambda$-CHOICE}
\newcommand{\INTLMC}{$\lambda\mu$-CHOICE}
\newcommand{\hpq}{\mathcal{Q}_{h}}

\newcommand{\lz}{\lambda(q)}
\newcommand{\mz}{\mu(q)}
\newcommand{\Rg}{\mathcal{R}_{\gamma}}
\newcommand{\lambdao}{\lambda_{1}}
\newcommand{\lo}{\lambda_{1}}
\newcommand{\lt}{\lambda_{2}}
\newcommand{\mo}{\mu_{1}}
\newcommand{\mt}{\mu_{2}}
\newcommand{\lambdat}{\lambda_{2}}
\newcommand{\muo}{\mu_{1}}
\newcommand{\mut}{\mu_{2}}
\newcommand{\newl}{u^{-1}(u_{c})}
\newcommand{\epv}{\epsilon_{V}}
\newcommand{\zms}{q_{\mu^*}}

\DeclareMathOperator*{\mini}{minimize}

\newcommand*{\qeda}{\hfill\ensuremath{\blacksquare}}

\begin{document}

\title{On the optimal tradeoff of average service cost rate, \\average utility rate, and average delay \\for the state dependent M/M/1 queue}

\author{Vineeth B. S. and Utpal Mukherji \\ Department of Electrical Communication Engineering,\\ Indian Institute of Science, Bangalore - 560012.\\\{vineeth,utpal\}@ece.iisc.ernet.in}

\date{}

\maketitle

\begin{abstract}
The optimal tradeoff between average service cost rate, average utility rate, and average delay is addressed for a state dependent M/M/1 queueing model, with controllable queue length dependent service rates and arrival rates.
For a model with a constant arrival rate $\lambda$ for all queue lengths, we obtain an asymptotic characterization of the minimum average delay, when the average service cost rate is a small positive quantity, $V$, more than the minimum average service cost rate required for queue stability.
We show that depending on the value of the arrival rate $\lambda$, the assumed service cost rate function, and the possible values of the service rates, the minimum average delay either: a) increases only to a finite value, b) increases without bound as $\log\nfrac{1}{V}$, c) increases without bound as $\frac{1}{V}$, or d) increases without bound as $\frac{1}{\sqrt{V}}$, when $V \downarrow 0$.
We then extend our analysis to a complementary problem, where the tradeoff of average utility rate and average delay is analysed for a M/M/1 queueing model, with controllable queue length dependent arrival rates, but a constant service rate $\mu$ for all queue lengths.
We obtain an asymptotic characterization of the average delay, when the average utility rate is a small positive $V$ less than the maximum utility rate which can be obtained with the queue being stable.
However, we find that in this case the minimum average delay always increases without bound, either as $\log\nfrac{1}{V}$, $\frac{1}{V}$, or $\frac{1}{\sqrt{V}}$ depending on $\mu$, the assumed utility rate function, and the possible values of the arrival rates.
We then consider a M/M/1 queueing model, with controllable queue length dependent service rates and arrival rates, for which we obtain an asymptotic characterization of the minimum average delay under constraints on both the average service cost rate as well as the average utility rate.
The results that we obtain are useful in obtaining intuition as well guidance for the derivation of similar asymptotic lower bounds, such as the Berry-Gallager asymptotic lower bound, for discrete time queueing models.
\end{abstract}

\section{Introduction}

We consider the tradeoff between average service cost rate, average utility rate, and average delay for a continuous time queueing model in this paper.
The mathematical model considered captures the problem of how a constrained/scarce resource should be dynamically allocated to randomly arriving demands, which may be subjected to admission control.
Herein, this dynamic allocation problem is modelled using a simple state dependent M/M/1 queueing model. 
Our primary motivation for modelling and studying this tradeoff problem as such, is to understand the tradeoff problems that arise in resource allocation problems in wireless networks.
The observations that can be obtained from studying this model also help in the study of discrete time queueing models (such as in \cite{vineeth_ncc13_dt} and \cite{vineeth_thesis}) which may be more appropriate in modelling wireless networks.

The state dependent M/M/1 model that we consider in this paper is a birth death process with the state corresponding to the queue length. 
We assume that the arrival rate and service rate at queue length $q$ are $\lambda(q)$ and $\mu(q)$ respectively.
We also assume that the arrival rates and service rates are controllable, i.e., for every $q \geq 0$, $\lambda(q)$ can be chosen from a set $\mathcal{X}_{\lambda}$ and for every $q > 0$, $\mu(q)$ can be chosen from a set $\mathcal{X}_{\mu}$.
A policy $\gamma$ is the choice of $\lambda(q)$ and $\mu(q), \forall q$.
We consider the problem of the choosing the optimal policy such that the average queue length is minimized subject to constraints on the average service cost rate and average utility rate.
Then, from Little's law, we obtain bounds on the minimum average delay subject to average service cost and utility rate constraints.

The average queue length $\Qg$ for a particular policy $\gamma$ is the time average of the expectation of the queue length $Q(t)$ (as in \cite{ross}), where $Q(t)$ is the state of the birth death process at time $t$ under $\gamma$.
We associate an utility rate function $u(.)$ with the arrival of customers and a cost rate function $c(.)$ with their service.
The utility rate function models the benefit in serving customers, while the cost rate function models the cost incurred in serving customers.
We assume that utility is accrued at the rate of $u(\lambda(Q(t)))$ and service cost is incurred at the rate of $c(\mu(Q(t)))$, where $u(\lambda)$ is a non-decreasing concave function of $\lambda$ and $c(\mu)$ is a non-decreasing convex function of $\mu$.
For the policy $\gamma$, the average utility rate $\Ug$ and average service cost rate $\Cg$ are defined as the time averages of the expectation of the utility rate $u(\lambda(Q(t)))$ and the expectation of the service cost rate $c(\mu(Q(t)))$ respectively.
The tradeoff problem that we consider is the minimization of $\Qg$, subject to a positive lower bound constraint $u_{c}$ on $\Ug$, and a positive upper bound constraint $c_{c}$ on $\Cg$, over all policies $\gamma$, i.e.,
\begin{eqnarray}
  \mini_{\gamma} & & \Qg, \nonumber \\
  \text{ such that } & & \Ug \geq u_{c} \text{ and } \Cg \leq c_{c}.
  \label{introduction:eq:mm1_genTradeoff}
\end{eqnarray}
In the following, we consider the above problem for the subset of \emph{admissible} policies.
Admissible policies are defined to be monotone, i.e., $\lambda(q)$ and $\mu(q)$ are non-increasing and non-decreasing functions of $q$ respectively.
This monotonicity property of admissible policies is motivated by the analysis of the unconstrained Lagrange dual problem, with the associated Lagrange dual function:
\begin{eqnarray}
  \mini_{\gamma} & & \Qg + \beta_{1} (\Cg - c_{c}) - \beta_{2} (\Ug - u_{c}),
  \label{introduction:eq:mm1_genTradeoff_dual}
\end{eqnarray}
where $\beta_{1}$ and $\beta_{2}$ are non-negative Lagrange multipliers.
We note that variations of the above problem have been formulated as Markov decision problems (MDP) and studied in \cite{george} and \cite{ata}.
We note that in this paper, we obtain bounds on the optimal solution of \eqref{introduction:eq:mm1_genTradeoff} in an asymptotic regime $\Re$, in which the constraint $c_{c}$ approaches a minimum average service cost rate (a function of $u_{c}$) which must be incurred for the queue to be stable.

\begin{figure}[!ht]
  \centering
  \includegraphics[width=\jImageWidth,height=\jImageHeight]{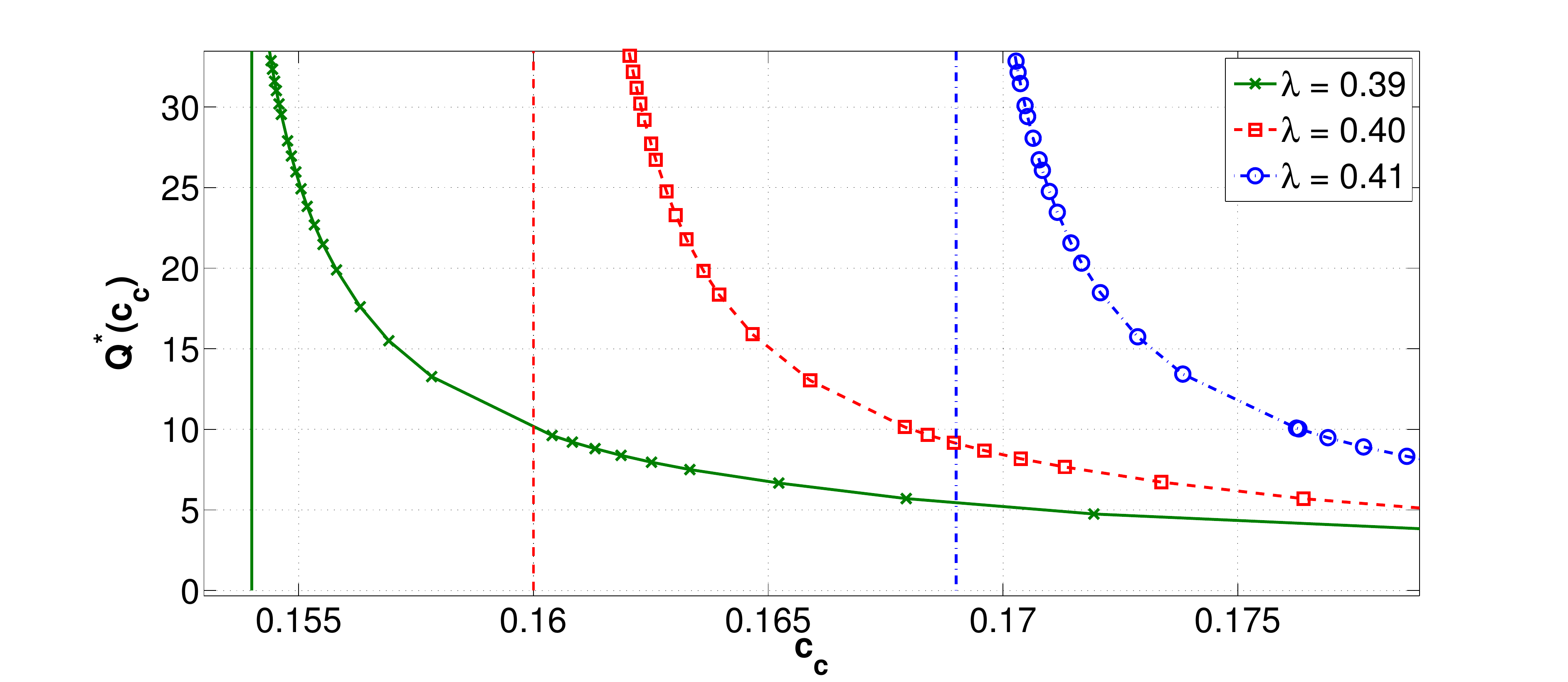}
  \caption{The optimal tradeoff curve with $\mu(q) \in \mathcal{S} = \brac{0, 0.2, 0.4, 0.5, 0.6, 0.8, 1}, \forall q$, service cost rate $c(\mu) = \mu^{2}, \forall \mu \in \mathcal{S}$, and $\lambda = 0.39, 0.40$, and $0.41$. The minimum average service cost rates are $\underline{c}(0.39) = 0.154, \underline{c}(0.40) = 0.160$, and $\underline{c}(0.41) = 0.169$.}
  \label{introduction:example}
\end{figure}

An example of the solution to \eqref{introduction:eq:mm1_genTradeoff} is shown in Figure \ref{introduction:example}, where we have considered a particular case with $\lambda(q)$ fixed to be $\lambda$ such that $u(\lambda) = u_{c}$.
The set of possible values that $\mu(q)$ can take is restricted to $\mathcal{S} = \brac{0, 0.2, 0.4, 0.5, 0.6, 0.8, 1}$.
The service cost constraint approaches the value $\underline{c}(\lambda)$ in the asymptotic regime $\Re$, where $\underline{c}(\mu):[0,1] \rightarrow \sR$ is the piecewise linear lower convex envelope of $c(\mu):S \rightarrow \sR$.
We assume that $c(\mu) = \mu^{2}, \mu \in \mathcal{S}$.
We have plotted the optimal solution
\footnote{For each $\lambda$, each point in the tradeoff curve $Q^*(c_{c})$ in Figure \ref{introduction:example} is obtained by numerically solving \eqref{introduction:eq:mm1_genTradeoff_dual} (via policy iteration) for a particular value of $\beta_{1}$ (since $u(\lambda) = u_{c}$, the second constraint is satisfied). From \cite{ma}, we have that any optimal policy $\gamma^*(\beta_{1})$ for \eqref{introduction:eq:mm1_genTradeoff_dual} is also optimal for \eqref{introduction:eq:mm1_genTradeoff} with $c_{c} = \overline{C}(\gamma^*(\beta_{1}))$. For a particular value of $\beta_{1}$, we obtain points $\brac{\overline{C}(\gamma^*(\beta_{1})), \overline{Q}(\gamma^*(\beta_{1}))}$.}
$Q^*(c_{c})$ as a function of $c_{c}$, for $\lambda = 0.39, 0.40$, and $0.41$.
We observe that the behaviour of $Q^*(c_{c})$ is dependent on the value of $\lambda$.
In this paper, we show that for $\lambda = 0.39$ and $0.41$, $Q^*(c_{c}) = \Theta\brap{\log\nfrac{1}{c_{c} - \underline{c}(\lambda)}}$ while for $\lambda = 0.40$ we show that $Q^*(c_{c})$ is $\Theta\nfrac{1}{c_{c} - \underline{c}(\lambda)}$.

\subsection{Related work}
\label{sec:relatedwork}
The tradeoff problem for continuous time single server queueing models has been addressed by many researchers.
The most common approach has been to pose this problem as a constrained Markov decision problem (CMDP) \cite{sennott} (as in \eqref{introduction:eq:mm1_genTradeoff}) which is then converted into a Markov decision problem (MDP) by a Lagrange relaxation \cite{ma} (as in \eqref{introduction:eq:mm1_genTradeoff_dual}).
The approach is then to find the optimal policy, or the optimal service rate and arrival rate as a function of queue length for \eqref{introduction:eq:mm1_genTradeoff_dual}.
Characterization of optimal policies for \eqref{introduction:eq:mm1_genTradeoff_dual} using the MDP approach yields structural properties, which are useful in reducing the search space for optimal policies, for example see \cite{weber}, \cite{george}, \cite{ata_pcstatic} and \cite{atamm1}.
In most cases, a monotonicity property of any optimal policy for \eqref{introduction:eq:mm1_genTradeoff_dual} is obtained, i.e., the optimal rate of service is a non-decreasing function of the queue length and the optimal arrival rate is a non-increasing function of the queue length.
Stidham and Weber \cite{weber} show that the optimal service rate is non-decreasing, and that the optimal arrival rate is non-increasing, for a state dependent M/G/1 model, where the objective is to minimize the expected total cost from any initial queue length by serving customers until the queue length is zero.
George and Harrison \cite{george} show that the optimal service rate is non-decreasing, for a state dependent M/M/1 model with $\lambda(q) = 1, \forall q$, where the objective is to minimize the time average cost, which is only composed of the service cost and the queue length.
Similar results have also been obtained by Ata in \cite{ata_pcstatic}, and Ata and Shneorson in \cite{atamm1}.
Surveys of the above approach can be obtained from \cite{koole} and \cite{sennott_text}.
The above monotonicity property is used to motivate the definition of admissible policies in the next section.
The complete characterization of an optimal arrival rate control policy of threshold form for a continuous time queueing model was obtained in \cite{perkins}.
An approximate solution to the tradeoff problem has been obtained by Ata et al. \cite{ata} by approximating the evolution of the number of the customers in the queue by a diffusion process, which enables the authors to find an  optimal stationary policy in closed form.

In this paper, we present an asymptotic analysis of the tradeoff between average service cost rate, average utility rate, and average queue length, for admissible policies, in the regime $\Re$.
The asymptotic analysis that we present for admissible policies is motivated by the asymptotic bounds derived by Berry and Gallager \cite{berry} for the tradeoff of average delay with average service cost for a discrete time queueing model.
For infinite buffer systems, it is intuitive that a minimum average service cost rate (a quantity such as $c(\lambda)$) has to be incurred to keep the queueing system stable.
Let $V$ be the difference between the average service cost and $c(\lambda)$.
Berry and Gallager show that the minimum average delay is at least $\Omega{\nfrac{1}{\sqrt{V}}}$ for any sequence of policies with average service cost $V$ more than $c(\lambda)$, for a discrete time queueing model\footnote{See Section \ref{sec:notation_conventions} for the definitions of $\Omega(.)$ and $\mathcal{O}(.)$}.
A sequence of policies is said to be order-optimal, if it achieves the optimal growth rate of the minimum average length with respect to $V$, in the asymptotic order sense.
For example, for the Berry Gallager model, if the average queue length of a sequence of policies increases as $\mathcal{O}\nfrac{1}{\sqrt{V}}$ with the average service cost being $V$ more than $c(\lambda)$, then the sequence of policies is order-optimal.
We note that families of \emph{good} policies can be identified by using the criterion of order-optimality.
The notion of order optimality for discrete time queueing models has been further explored in many other papers, such as \cite{neely_mac}, \cite{chaporkar}, and surveyed comprehensively in \cite{neely}.
Order optimality has also been explored for discrete time finite buffer queueing models.
In \cite[Chapter 6]{berry_thesis} it is shown that for a finite buffer discrete time queueing model, as the buffer size $B$ goes to infinity, for any sequence of policies such that the buffer overflow probability is $o\brap{\frac{1}{B^{2}}}$, the average service cost is at least $\Omega\brap{\frac{1}{B^{2}}}$ more than $c_{min}$.
We note that an order optimality result for a two sided birth death queueing model was obtained by Ramaiyan \cite{venkithesis}.
Our approach which uses the detailed balance equations to solve for the stationary distribution of the birth-death process is similar to that in \cite{venkithesis}.
However, we obtain observations about the nature of the stationary distribution which were previously not available.

Order optimality results are not available for continuous time queueing models, which motivates us to develop asymptotic upper and lower bounds for the continuous time M/M/1 model considered in this paper.
Furthermore, we shall see that the analysis which leads to these bounds leads to observations about the stationary probability distribution for the M/M/1 model, which can be used in deriving new asymptotic lower bounds for discrete time queueing models as in \cite{vineeth_thesis} and \cite{vineeth_ncc13_dt}.

\subsection{Overview and contributions}
The state dependent M/M/1 model is discussed in Section \ref{sec:system_model}. 
Assumptions on the properties of the utility and cost functions as well as the average queue length, average utility rate, and average service cost rate are also defined in the same section.
The tradeoff problem is then set up in Section \ref{sec:problem_formulation}. We define the restricted set of admissible policies in this section.
This restriction is motivated by the monotonicity properties of optimal policies, as reviewed in the previous section.
We also present necessary and sufficient conditions for the feasibility of the tradeoff problem in the same section.
We then define three cases of the tradeoff problem: \INTMC, \INTLC, and \INTLMC\, which are then analysed in the rest of the paper.
We note that \INTMC, \INTLC, and \INTLMC\, are queueing models with service rate control and no arrival rate control, no service rate control and arrival rate control, and both service and arrival rate control, respectively.
Thus, they cover all cases of interest.
We also define the asymptotic regime $\Re$ in Section \ref{sec:problem_formulation}.
In Sections \ref{sec:problemp2}, \ref{sec:intlc_analysis}, and \ref{sec:problemp3} we analyse \INTMC, \INTLC, and \INTLMC\, respectively.
We note that results for a special case of \INTMC, with $\mathcal{X}_{\mu}$ being a discrete set, were presented in \cite{vineeth_ncc13_ct}.
We conclude the paper in Section \ref{sec:summary}.

\subsection{Notation and conventions}
\label{sec:notation_conventions}
We use the following notation for the asymptotic bounds: (i) $f(x)$ is $\mathcal{O}(g(x))$ if there exists a $c > 0$ such that $\lim_{x \rightarrow 0} \frac{f(x)}{g(x)} \leq c$; $f(x), g(x) \geq 0$, (ii) $f(x)$ is $\Omega(g(x))$ if there exists a $c > 0$ such that $\lim_{x \rightarrow 0} \frac{f(x)}{g(x)} \geq c$; $f(x), g(x) \geq 0$, (iii) $f(x)$ is $o(g(x))$ if $\lim_{x \downarrow 0} \frac{f(x)}{g(x)} = 0$, and (iv) $f(x)$ is $\omega(g(x))$ if $\lim_{x \downarrow 0} \frac{f(x)}{g(x)} = \infty$.
All logarithms are natural logarithms unless specified otherwise.
Sequences which are monotonically increasing to a limit point are denoted using $\uparrow$, while those monotonically decreasing are denoted using $\downarrow$.
The stationary version of a random process is denoted by dropping the time index, e.g. $Q \sim Q(t)$.
We denote the set of non-negative integers and non-negative real numbers by $\sZ$ and $\sR$ respectively.

\section{System model}
\label{sec:system_model}

\begin{figure}
  \centering
  \includegraphics[width=\jImageWidth,height=\jImageHeight]{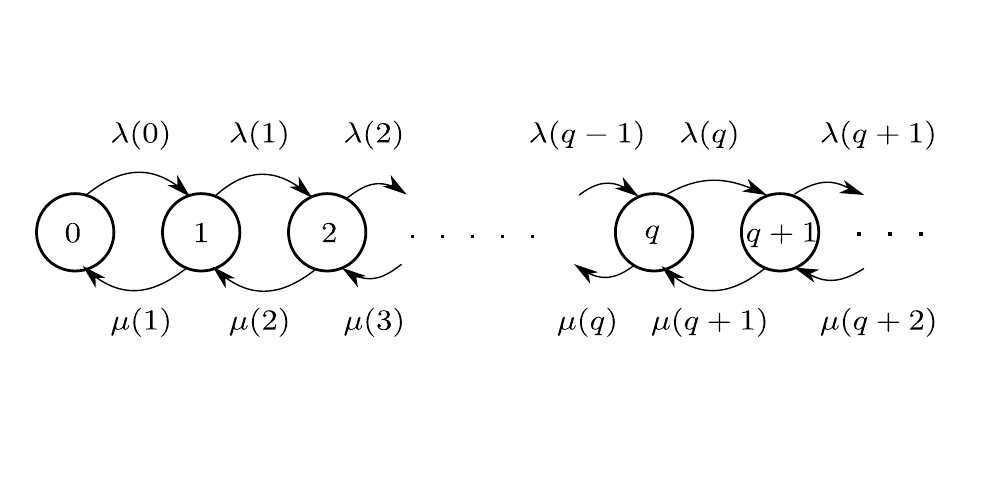}
  \vspace{-0.5in}
  \caption{The birth death process under a particular policy $\gamma$}
  \label{fig:mm1model}
\end{figure}

The system evolves in continuous time, which is denoted by $t \in \mathbb{R}_{+}$.
The number of customers in the queue at time $t$ (including the one in service, if any)  is denoted by $Q(t) \in \mathbb{Z}_{+}$.
The state dependent M/M/1 model for the process $Q(t)$ is a birth death process with birth rate $\lambda(q) \in \mathcal{X}_{\lambda}$ from $q$ to $q + 1$ for all $q \in \sZ$, and death rate $\mu(q) \in \mathcal{X}_{\mu}$ from $q$ to $q - 1$ for all positive $q$.
A policy $\gamma$ is the sequence $(\mu(0) = 0, \lambda(0), \mu(1), \lambda(1) \cdots)$
\footnote{We note that for the tradeoff problems that we are interested in, if we restrict to policies that update the control of arrival rate and service rate only at customer arrivals and/or departures, then as in \cite{hernandez}, it can be shown that we can restrict to stationary policies.}.
The state transition diagram of the birth-death process for a policy $\gamma$ is shown in Figure \ref{fig:mm1model}.
The set of all policies is denoted as $\Gamma$.

We recall that utility is accrued at the rate of $u(\lambda(Q(t)))$ and cost is incurred at the rate of $c(\mu(Q(t)))$ at time $t$.
The functions $u(.)$ and $c(.)$ are assumed to satisfy the following properties:
\begin{description}
\item[U1 :]{The function $u(\lambda) : \mathcal{X}_{\lambda} \rightarrow \mathbb{R}_{+}$ is strictly increasing and concave in $\lambda$, with $u(0) = 0$. The set of all possible arrival rates is $\mathcal{X}_{\lambda}$.}
\item[C1 :]{The function $c(\mu) : \mathcal{X}_{\mu} \rightarrow \mathbb{R}_{+}$ is strictly increasing and convex in $\mu$, with $c(0) = 0$.
The set of all possible service rates is $\mathcal{X}_{\mu}$.}
\end{description}
We assume that $\mathcal{X}_{\lambda} = [r_{a,min}, r_{a,max}]$ and $\mathcal{X}_{\mu} = [r_{min}, r_{max}]$, where $r_{a,min} < r_{max}$.
Let $u^{-1} : \mathbb{R}_{+} \rightarrow \mathcal{X}_{\lambda}$ and $c^{-1} : \mathbb{R}_{+} \rightarrow \mathcal{X}_{\mu}$ be the inverse functions of $u(.)$ and $c(.)$ respectively.

The average service cost rate for the policy $\gamma$, $\overline{C}(\gamma)$ is defined as
\begin{equation}
  \overline{C}(\gamma) = \lim_{T \rightarrow \infty}  \frac{1}{T} \mathbb{E} \bras{\int_{0}^{T} c(\mu(Q(t)))dt \middle \vert Q(0) = q_{0}}.
  \label{eq:sysmodel:sercostdef}
\end{equation}
The average utility rate for the policy $\gamma$, $\overline{U}(\gamma)$ is defined as
\begin{equation}
  \overline{U}(\gamma) = \lim_{T \rightarrow \infty} \frac{1}{T} \mathbb{E} \bras{\int_{0}^{T} u(\lambda(Q(t)))dt \middle \vert Q(0) = q_{0}}.
  \label{eq:sysmodel:utilitydef}
\end{equation}
The average queue length for the policy $\gamma$, $\overline{Q}(\gamma)$ is defined as
\begin{equation}
  \overline{Q}(\gamma) = \lim_{T \rightarrow \infty} \frac{1}{T} \mathbb{E} \bras{\int_{0}^{T} Q(t)dt \middle \vert Q(0) = q_{0}}.
  \label{eq:sysmodel:qlength}
\end{equation}
In this paper, we restrict attention to policies for which the above three performance measures are independent of the initial state $q_{0}$, hence in the above definitions the dependence of these quantities on $q_{0}$ is not made explicit.
We note that the above definition of average utility rate (as in \cite{atamm1}) is much more general and encompasses scenarios where the utility of average throughput is of interest (e.g. as in \cite{neely_utility}).

\section{Problem formulation}
\label{sec:problem_formulation}

In this paper, we consider the tradeoff problem \eqref{introduction:eq:mm1_genTradeoff} for a set of admissible policies only.
The set of admissible policies is defined as follows.

\begin{definition}{(Stability)}
A policy $\gamma$ is defined to be stable if: (i) the birth death process $Q(t)$ under policy $\gamma$ has a single positive recurrent class $\mathcal{R}_{\gamma}$, and stationary distribution $\pi_{\gamma}$ and (ii) the expected cumulative (i) queue cost, (ii) service cost, and (iii) utility until, $\mathcal{R}_{\gamma}$ is hit starting from any initial state $q_{0}$, are finite.
\end{definition}

\begin{definition}(Admissibility)
A policy $\gamma$ is admissible, if: (G1) it is stable, (G2) the sequence $(\mu(0), \mu(1), \mu(2), \cdots)$ is non-decreasing, and (G3) the sequence $(\lambda(0), \lambda(1), \lambda(2), \cdots)$ is non-increasing.
We define the set of admissible policies as
\begin{equation*}
  \Gamma_{a} \Deq \{ \gamma : \gamma \in {\Gamma}, \gamma \text{ is admissible}\}.
\end{equation*}
\end{definition}
\begin{remark}
  We restrict to the set of admissible policies, since there exists an optimal policy for \eqref{introduction:eq:mm1_genTradeoff} which possesses the properties G1, G2, and G3 in many cases.
  Consider the unconstrained MDP in \eqref{introduction:eq:mm1_genTradeoff_dual} denoted as $MDP(\beta_{1}, \beta_{2})$.
  We note that this MDP is obtained by uniformizing $Q(t)$ at rate $r_{u}$ with single stage cost $\frac{q + \beta_{1} c(\mu) - \beta_{2} u(\lambda)}{r_{u}}$ as in \cite{atamm1}.
  Then from \cite{atamm1} we know that an optimal policy $\gamma^*(\beta_{1}, \beta_{2}) \in \Gamma_{a}$ exists for $MDP(\beta_{1}, \beta_{2})$.
  From \cite{ma}, we have that $\gamma^*(\beta_{1}, \beta_{2})$ is also optimal for \eqref{introduction:eq:mm1_genTradeoff} if $c_{c} = \overline{C}(\gamma^*(\beta_{1}, \beta_{2}))$ and $u_{c} = \overline{U}(\gamma^*(\beta_{1}, \beta_{2}))$.
  Let $\Gamma^*(\beta_{1}, \beta_{2})$ be the set of all optimal admissible policies for $MDP(\beta_{1}, \beta_{2})$.
  Let $\mathcal{O}^{d}_{u} = \brac{(\overline{C}(\gamma^*(\beta_{1}, \beta_{2})), \overline{U}(\gamma^*(\beta_{1}, \beta_{2}))), \gamma^*(\beta_{1}, \beta_{2}) \in \Gamma^*(\beta_{1}, \beta_{2}), \forall \beta_{1}, \beta_{2} \geq 0}$.
  Then, there exists an optimal $\gamma \in \Gamma_{a}$ for the constrained optimization problem \eqref{introduction:eq:mm1_genTradeoff}, when $(c_{c}, u_{c}) \in \mathcal{O}^{d}_{u}$.
  \label{remark:definitionOdu}
\end{remark}

We note that for any $\gamma \in \Gamma_{a}$, we have that 
\begin{eqnarray*}
  \overline{C}(\gamma) & = & \mathbb{E}_{\pi_{\gamma}} c(\mu(Q)), \\
  \overline{U}(\gamma) & = & \mathbb{E}_{\pi_{\gamma}} u(\lambda(Q)), \text{ and } \\
  \overline{Q}(\gamma) & = & \mathbb{E}_{\pi_{\gamma}} Q,
\end{eqnarray*}
where $Q \sim \pi_{\gamma}$.
We also note that for $\gamma \in \Gamma_{a}$ the performance measures are independent of the initial state $q_{0}$.
The set $\Rg$ for a $\gamma \in \Gamma_{a}$ is contiguous and is of the form $(q_{r,l}, \cdots, q_{r,u})$, where $\mu(q_{r,l}) = 0$ and $\lambda(q_{r,u}) = 0$.
Also, for $q > q_{r,l}, \mu(q) > 0$ and $q < q_{r,u}, \lambda(q) > 0$.

In the following discussion we consider the problem TRADEOFF,
\begin{eqnarray}
  \text{ minimize }_{\gamma \in \Gamma_{a}} & \overline{Q}(\gamma) \nonumber \\
  \text{ such that } & \overline{C}(\gamma) \leq c_{c}, \nonumber \\
  \text{ and } & \overline{U}(\gamma) \geq u_{c},
\end{eqnarray}
where we minimize over the set $\Gamma_{a}$ only.
The optimal value of TRADEOFF is denoted as $Q^*(c_{c}, u_{c})$.
\begin{remark}
  We note that if $(c_{c}, u_{c}) \in \mathcal{O}^{d}_{u}$ then: (i) from Remark \ref{remark:definitionOdu} we have that $Q^*(c_{c}, u_{c})$ is the optimal value of \eqref{introduction:eq:mm1_genTradeoff}, (ii) there exists an optimal admissible policy for TRADEOFF, which is also optimal for \eqref{introduction:eq:mm1_genTradeoff}.
  If $(c_{c}, u_{c}) \not \in \mathcal{O}^{d}_{u}$ then we note that (i) and (ii) above may not hold.
  In Section \ref{sec:discussion} we discuss how TRADEOFF can be used to obtain solutions for \eqref{introduction:eq:mm1_genTradeoff} if $(c_{c}, u_{c}) \not \in \mathcal{O}^{d}_{u}$.
\end{remark}

We now address the question of feasibility for TRADEOFF.
Consider the example where $\XL = \brac{\lambda}$, then we note that the average service rate $\Exp_{\pig} \mu(Q)$ has to be $\lambda$ for any $\gamma \in \Gamma_{a}$.
Therefore, $\Cg$ has to be at least $c(\lambda)$ (this arises from Jensen's inequality applied to $\Exp_{\pig} c(\mu(Q))$.
Thus, we have that for a given $c_{c}$, if there exists any feasible policy for TRADEOFF then $c_{c} \geq c(\lambda)$.
The following lemma gives the necessary condition for feasibility of TRADEOFF for more general cases.
\begin{lemma}
  If TRADEOFF has any feasible solutions, then $u^{-1}(u_{c}) \leq c^{-1}(c_{c})$.
  \label{lemma:feasible_solns}
\end{lemma}
The proof is given in Appendix \ref{appendix:proof:lemma:feasible_solns}.
For TRADEOFF, if $c^{-1}(c_{c}) > u^{-1}(u_{c})$, then it can be shown that a feasible admissible policy exists.
We present this feasible admissible policy $\gamma \in \Gamma_{a}$ on a case by case basis in the following discussion.

We also note that if even $c^{-1}(c_{c}) > u^{-1}(u_{c})$, unless $(c_{c}, u_{c}) \in \mathcal{O}^{d}_{u}$, an optimal policy $\gamma^* \in \Gamma_{a}$ may not exist for TRADEOFF.
This leads to the following definition of $\epsilon$-optimal policies which are \emph{close} to optimal.

\begin{definition}{($\epsilon$-optimal policy)}
  Suppose $c^{-1}(c_{c}) > u^{-1}(u_{c})$.
  Then there exists some feasible $\gamma \in \Gamma_{a}$ for TRADEOFF.
  By definition, for every $\epsilon > 0$, there exists a feasible $\gamma_{\epsilon}$ such that $\overline{Q}(\gamma_{\epsilon}) < Q^*(c_{c}, u_{c}) + \epsilon$. 
  Then, $\gamma_{\epsilon}$ is defined to be $\epsilon$-optimal for $(c_{c}, u_{c})$.
\end{definition}
We note that TRADEOFF is feasible iff $c_{c} > c(u^{-1}(u_{c}))$.
So $c(u^{-1}(u_{c}))$ can be interpreted as the infimum of the average service cost rate which must be incurred for TRADEOFF to be feasible.
We analyse TRADEOFF in an asymptotic regime $\Re$ defined as follows.
\begin{definition}{(The asymptotic regime $\Re$)}
We consider a non-increasing sequence $c_{c,k}$ and a non-decreasing sequence $u_{c,k}$.
The asymptotic regime $\Re$ is defined as the regime in which $c_{c,k} - c(u^{-1}(u_{c,k})) \downarrow 0$.
\end{definition}

In this paper, we obtain bounds on the optimal value of TRADEOFF for three different cases: \INTMC, \INTLC, and \INTLMC\, which are defined below.
\begin{definition}{(\INTMC)}
We consider the case with $\XL = \brac{\lambda}$ and $\XM = [0, r_{max}]$.
Then for $\gamma$, $\Ug = u(\lambda)$. 
We assume that $\lambda$ is such that $u(\lambda) = u_{c}$ in TRADEOFF. 
We note that only the service rate can be controlled for this case.
We note that in the asymptotic regime $\Re$ for \INTMC\, $c_{c,k} \downarrow c(\lambda)$ since $\lambda$ is fixed to be $u^{-1}(u_{c})$.
\end{definition}

\begin{definition}{(\INTLC)}
We consider the case with $\XM = \brac{\mu}$ and $\XL = [r_{a,min}, r_{a,max}]$. 
Then for $\gamma$, $\Cg \leq c(\mu)$. 
We assume that $\mu$ is such that $c(\mu) = c_{c}$ in TRADEOFF. 
We note that only the arrival rate can be controlled for this case.
We note that in the asymptotic regime $\Re$ for \INTMC\, $u_{c, k} \uparrow u(\mu)$ since $\mu$ is fixed to be $c^{-1}(c_{c})$.
\end{definition}

\begin{definition}{(\INTLMC)}
In this case, we assume that $\XM = [0, r_{max}]$ and $\XL = [r_{a,min}, r_{a,max}]$.
We note that in the asymptotic regime $\Re$ for \INTLMC, $c_{c,k} - c(u^{-1}(u_{c,k})) \downarrow 0$, where either one of $c_{c,k}$ or $u_{c,k}$ can be fixed.
\end{definition}

We are motivated to study these three cases since they correspond to the study of queueing systems with only service rate control, only  admission control, and both service rate and admission control respectively.
In the following discussion, we assume that $c(\mu)$ is convex and either strictly convex or piecewise linear and $u(\lambda)$ is concave and either strictly concave or piecewise linear.
In practice service costs have increasing marginal returns, while utilities have diminishing marginal returns, which leads to the convexity and concavity assumptions for $c(\mu)$ and $u(\lambda)$ respectively.
However, the strict or piecewise linear nature of $c(\mu)$ or $u(\lambda)$ are motivated by scenarios which arise for discrete time queueing models, such as those considered in \cite[Chapter 4]{vineeth_thesis}.
We note that for discrete time queueing models, the possible control variables are the batch size ($A(q)$) of the number of customers admitted and the batch size ($S(q)$) of the number of customers which are served in each slot, which are possibly random functions of the queue length $q$.
Since for the M/M/1 model the control variables $\lambda(q)$ and $\mu(q)$ are deterministic, one way to capture the nature of the control variables for the discrete time model via the M/M/1 model is to have a correspondence between $\lambda(q)$ and $\Exp A(q)$, and $\mu(q)$ and $\Exp S(q)$.
We note that the service cost rate and utility rate for discrete time models are $\Exp u_{d}(A(q))$ and $\Exp c_{d}(S(q))$ respectively, where $u_{d}$ and $c_{d}$ are the utility and cost functions for the discrete time model.
The maximum value of $\Exp u_{d}(A(q))$ is $\overline{u}(\Exp A(q))$, where $\overline{u}$ is the upper concave envelope of $u_{d}$, and the minimum value of $\Exp c_{d}(S(q))$ is $\underline{c}(\Exp S(q))$, where $\underline{c}$ is the lower convex envelope of $c_{d}$.
Again, one way in which to capture the nature of the cost and utility functions for the discrete time model via the M/M/1 model is to have a correspondence between $u(\lambda)$ and $\overline{u}(\Exp A(q))$, and $c(\mu)$ and $\underline{c}(\Exp S(q))$.
Depending upon whether $S(q)$ (or $A(q)$) takes values in $\sR$ or $\sZ$ the form of the function $c(.)$ (or $u(.)$) can be strictly convex or piecewise linear.
These choices turn out to be good in retrospect.

\section{Analysis of \INTMC}
\label{sec:problemp2}

\begin{figure}[h]
  \centering
  \includegraphics[width=160mm,height=80mm]{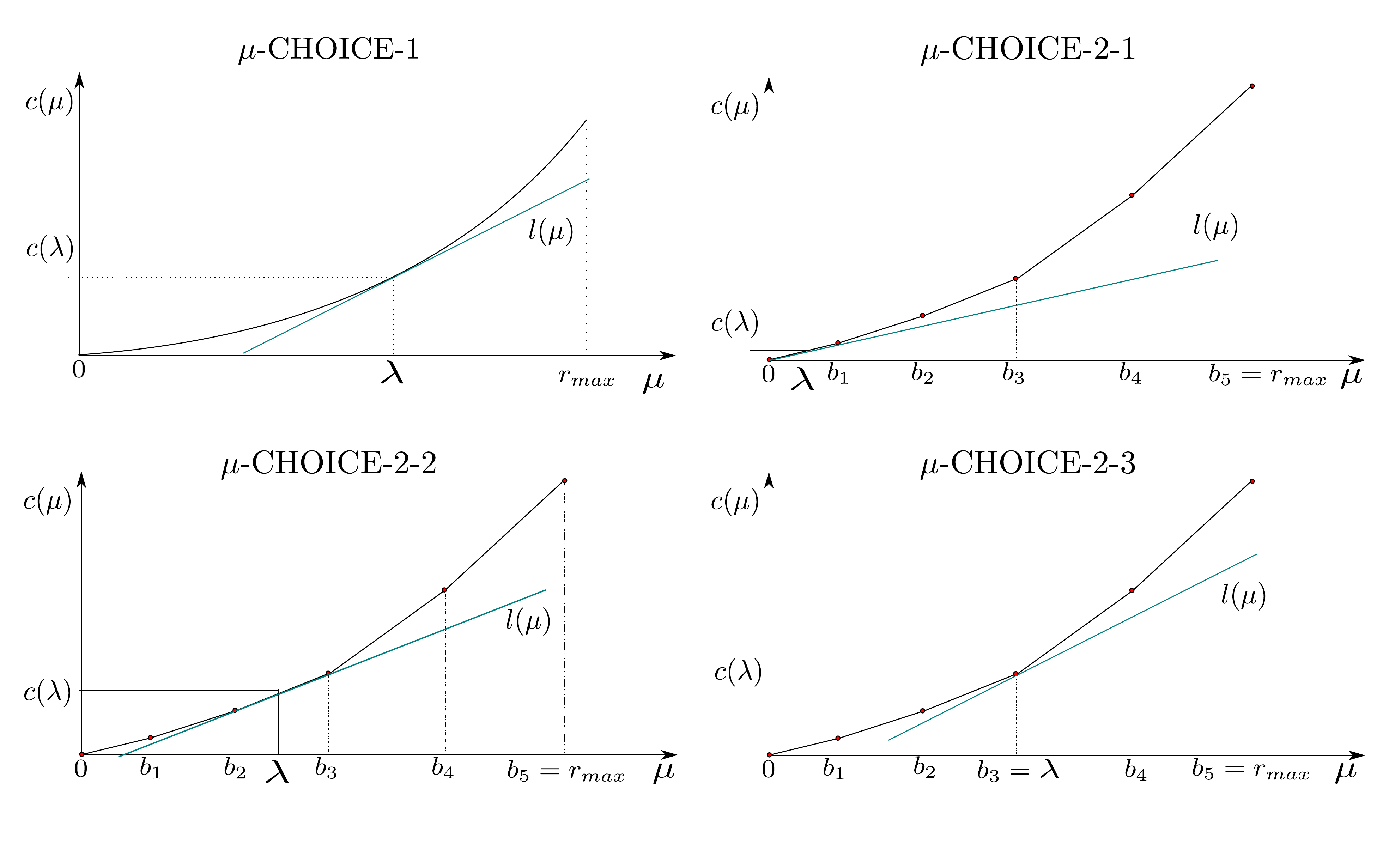}
  \caption{Illustration of the relationship between $\lambda$, $\mu_l$, and $\mu_{u}$ along with the minimum average cost $c(\lambda)$ and the line $l(\mu)$ for the four cases of the \INTMC\, problem.}
  \label{fig:p2servicecost_convexhull}
\end{figure}

We note that for \INTMC, $\lambda(q) = \lambda$ and $\mu(q) \in [0,r_{max}]$, $\forall q \in \mathbb{Z}_{+}$.
The tradeoff problem for \INTMC\, is
\begin{eqnarray*}
  \text{ minimize }_{\gamma \in \Gamma_{a}} & \overline{Q}(\gamma) \nonumber \\
  \text{ such that } & \overline{C}(\gamma) \leq c_{c}. \nonumber
\end{eqnarray*}
The optimal value of \INTMC\, is denoted as $Q^*(c_{c})$.
The study of \INTMC\, is classified into:
\begin{description}
\item[\INTMC-1:] $c(\mu)$ is strictly convex for $\mu \in [0,r_{max}]$, and
\item[\INTMC-2:] $c(\mu)$ is piecewise linear, i.e., (a) there exists a minimal partition of $[0,r_{max}]$ into intervals $\{[a_{i},b_{i}], i \in \{1,\dots,P\}\}$ with $a_{1} = 0$, $b_{P} = r_{max}$, and $b_{i} = a_{i + 1}$, and (b) there are linear functions $f_{i}$ such that $\forall \mu \in [a_{i},b_{i}], f_{i}(\mu) = c(\mu)$.
\end{description}
It turns out that the asymptotic behaviour of $Q^*(c_{c})$ for \INTMC-2 depends on the behaviour of $c(\mu)$ in a neighbourhood of $\mu = \lambda$.
Therefore, we consider the following cases for \INTMC-2:
\begin{description}
\item[\INTMC-2-1:] $\lambda \in (0, b_{\lambda} \Deq b_{1} )$,
\item[\INTMC-2-2:] $\lambda \in (a_{\lambda} \Deq a_{i} ,b_{\lambda} \Deq b_{i})$ for some $i \in \{2,\dots,P\}$, and,
\item[\INTMC-2-3:] $a_{\lambda} \Deq \lambda = a_{i}$ for some $i \in \{2,\dots,P\}$.
\end{description}
The different cases are illustrated in Figure \ref{fig:p2servicecost_convexhull}.
We also define a line $l(\mu)$ as follows: (i) for \INTMC-1, $l(\mu)$ is defined as the tangent to $c(\mu)$ at $\mu = \lambda$, (ii) for \INTMC-2-1 and \INTMC-2-2, $l(\mu)$ is defined to be line through $(a_{\lambda}, c(a_{\lambda}))$ and $(b_{\lambda}, c(b_{\lambda}))$, and (iii) for \INTMC-2-3, $l(\mu)$ is defined as any line through $(a_{\lambda} = a_{i}, c(a_{\lambda}))$ with slope $m$ such that $\frac{c(a_{i}) - c(a_{i-1})}{a_{i} - a_{i - 1}} < m < \frac{c(a_{i + 1}) - c(a_{i})}{a_{i + 1} - a_{i}}$.

We note that for \INTMC-1, the function $c(\mu)$ is strictly convex for every $\mu \in [0, r_{max}]$ and therefore the asymptotic behaviour is the same for all $\lambda$.
We now present the asymptotic lower bounds for $Q^*(c_{c})$ in the regime $\Re$, in which $c_{c} \downarrow c(\lambda)$ for the above cases.
Then, in Section \ref{sec:intmc_ub}, we show that there exists a sequence of admissible policies $\gamma_{k}$ such that $\Cgk \downarrow c(\lambda)$.

\subsection{Asymptotic lower bounds}
\label{sec:intmc_lb}

We obtain an asymptotic lower bound on $\overline{Q}(\gamma)$ by: (a) obtaining an upper bound on the stationary probability for a certain set of service rates in terms of $\overline{C}(\gamma)$ and $c(\lambda)$, (b) relating the stationary probability of this set of service rates to the stationary probability $\pi(q)$, of a set of queue lengths, and (c) obtaining a lower bound on $\overline{Q}(\gamma)$ in terms of $\pi(q)$.
We note that even though $\mu(q) \in [0, r_{max}]$, the set $\brac{\mu(q), q \in \mathbb{Z}_{+}}$ is only countable.
Let $\brap{\mu_{0} = 0, \dots, \mu_{k}, \dots}$, with $\mu_{k} < \mu_{k + 1}$ denote the set of service rates that is used by a policy $\gamma$.
Also let $\pi_{\mu}(k) = \sum_{q : \mu(q) = \mu_{k}} \pi(q)$ be the stationary probability of using rate $\mu_{k}$.
The relationships in steps (a) and (b) above are obtained in the following lemma.

\begin{lemma}
  Let $\pi$ be the stationary probability distribution of $Q(t)$ and let $Q \sim \pi$. Let $V \Deq c_{c} - c(\lambda)$. Let $S \subseteq [0, r_{max}]$ and $Q_{S} \Deq \brac{q : \mu(q) \in S}$. Then for any $\epsilon_{V} > 0$, 
  \begin{enumerate}
    \item for \INTMC-1 with $S = [0, \lambda - \epsilon_{V}) \bigcup (\lambda + \epsilon_{V}, r_{max}]$ we have that $Pr\brac{\mu(Q) \in S} \leq \frac{V}{a_{1} \epsilon_{V}^{2}}$, where $a_{1} > 0$,
    \item for \INTMC-2-1 and \INTMC-2-2 with $S = [0, a_{\lambda} - \epsilon_{V}) \bigcup (b_{\lambda} + \epsilon_{V}, r_{max}]$ we have that $Pr\brac{\mu(Q) \in S} \leq \frac{V}{m_{a} \epsilon_{V}}$,
    \item for \INTMC-2-3 with $S = [0, \lambda - \epsilon_{V}) \bigcup (\lambda + \epsilon_{V}, r_{max}]$ we have that $Pr\brac{\mu(Q) \in S} \leq \frac{V}{m_{a} \epsilon_{V}}$,
  \end{enumerate}
  where $m_{a} > 0$.
  Since $Pr\brac{Q \in Q_{S}} = Pr\brac{\mu(Q) \in S}$, the same bounds hold for $Pr\brac{Q \in Q_{S}}$.
  We also note that the above bounds are valid for any subset of $S$.
  \label{lemma:statprobbound}
\end{lemma}
The proof is given in Appendix \ref{appendix:proof:lemma:statprobbound}.
\begin{remark}
  We note that the set $\Rg$ is dependent on the policy $\gamma$.
  Since $\brac{0, \dots, q_{r,l} - 1}$ and $\brac{q_{r,u} + 1, \dots}$ are transient states, we note that $\Qg, \Cg$, and $\Ug$ can be obtained by considering a birth death process $Q_{r}(t)$ on $\brac{0, \dots, q_{r,u} - q_{r,l}}$.
  The process $Q_{r}(t)$ is obtained by restricting $Q(t)$ to $\brac{q_{r,l},\dots,q_{r,u}}$ and then relabelling the states.
  The birth rate $\lambda_{r}(q)$ and death rate $\mu_{r}(q)$ of $Q_{r}(t)$ are $\lambda(q + q_{r,l})$ and $\mu(q + q_{r,l})$, respectively.
  We have that
  \begin{eqnarray*}
    \Cg & = & \Exp c\brap{\mu(Q_{r} + q_{r,l})},\\
    \Ug & = & \Exp u\brap{\mu(Q_{r} + q_{r,l})},\\
    \Qg & = & q_{r,l} + \Exp Q_{r}.
  \end{eqnarray*}
  Since $q_{r,l} \geq 0$, $\Exp Q_{r} \leq \Qg$.
  In the following, we use $\Exp Q_{r}$ to obtain asymptotic lower bounds on $\Qg$.
  \label{remark:recurrentclass}
\end{remark}
We first consider \INTMC-1, for which $c(\mu)$ is a strictly convex function of $\mu \in [0, r_{max}]$.
We make the following assumption regarding $c(\mu)$ at $\mu = \lambda$.
\begin{description}
  \item[C2 :] For \INTMC-1, the second derivative of $c(\mu)$ is non-zero at $\mu = \lambda$.
\end{description}
The above assumption has been used in \cite{berry}.
We note that since $c(\mu)$ is strictly convex, the second derivative of $c(\mu)$ is non-zero for all $\mu \in [0, r_{max}]$ except for $\mu$ in a countable set.
In the derivation of the asymptotic lower bounds in this paper, we assume that the process $Q(t)$ is irreducible.
However, the assumption of irreducibility can be removed by considering the process $Q_{r}(t)$ instead of $Q(t)$ as defined in Remark \ref{remark:recurrentclass}.
\begin{lemma}
  For \INTMC-1, for any sequence of admissible policies $\gamma_{k}$ such that $\overline{C}(\gamma_{k}) - c(\lambda) = V_{k} \downarrow 0$ we have that $\overline{Q}(\gamma_{k}) = \Omega\nfrac{1}{\sqrt{V_{k}}}$.
  Therefore, $Q^*(c_{c}) = \Omega\nfrac{1}{\sqrt{c_{c} - c(\lambda)}}$.
  \label{prop:p21lb}
\end{lemma}
The proof is given in Appendix \ref{appendix:proof:prop:p21lb}.
We now obtain asymptotic lower bounds for \INTMC-2.
\begin{lemma}
  For \INTMC-2-1, for any sequence of admissible policies $\gamma_{k}$ with $\overline{C}(\gamma_{k}) - c(\lambda) = V_{k} \downarrow 0$ we have that $\overline{Q}(\gamma_{k}) = \frac{\lambda}{b_{\lambda} - \lambda} - \mathcal{O}\brap{V^{1 - \delta}\log\nfrac{1}{V}}$, for $0 < \delta < 1$.
  Therefore, $Q^*(c_{c}) = \frac{\lambda}{b_{\lambda} - \lambda} - \mathcal{O}\brap{(c_{c} - c(\lambda))^{1 - \delta}\log\nfrac{1}{c_{c} - c(\lambda)}}$, for $0 < \delta < 1$.
  \label{prop:p221lb}
\end{lemma}
The proof is given in Appendix \ref{appendix:proof:prop:p221lb}. 
We note that if $\mathcal{X}_{\mu}$ were a discrete set, as in the case of FINITE-$\mu$CHOICE-1 in \cite[Chapter 2]{vineeth_thesis} then an asymptotic order of $\mathcal{O}\brap{V\log\nfrac{1}{V}}$ can be obtained.
However, for \INTMC-2-1, we are only able to show that the order is $\mathcal{O}\brap{V^{1 - \delta}\log\nfrac{1}{V}}$, where $\delta$ can be made arbitrarily close to zero.

\begin{lemma}
  For \INTMC-2-2, for any sequence of admissible policies $\gamma_{k}$ with $\overline{C}(\gamma_{k}) - c(\lambda) = V_{k} \downarrow 0$, we have that $\overline{Q}(\gamma_{k}) = \Omega\brap{\log\nfrac{1}{V_{k}}}$.
  \label{prop:p222lb}
\end{lemma}
The proof is given in Appendix \ref{appendix:proof:prop:p222lb}.
\begin{remark}
    We note that the lower bounding technique in \cite{berry} and \cite{neely_utility} can be used to obtain the asymptotic lower bounds for \INTMC-1 and \INTMC-2-2 respectively.
  This method, which considers a uniformized version of $Q(t)$, is outlined in \cite[Chapter 3]{vineeth_thesis}.
  However, using the stationary probability of the queue length has its advantages, since it gives us additional insights into the form of the optimal policy.
  We use bounds on the stationary probability of the queue length to obtain an asymptotic characterization of the cardinality of the set $Q_{S}$, where $Q_{S} = \brac{q:\mu(q) \in S}$ (e.g. $S = [0, \lambda - V^{\frac{1}{2}}]$ for \INTMC-1) in \cite[Proposition 2.3.16, Lemma 3.2.18]{vineeth_thesis}.
\end{remark}
\begin{lemma}
  For \INTMC-2-3, for any sequence of admissible policies $\gamma_{k}$ with $\overline{C}(\gamma_{k}) - c(\lambda) = V_{k} \downarrow 0$, we have that $\overline{Q}(\gamma_{k}) = \Omega\nfrac{1}{V_{k}}$.
  \label{prop:p223lb}
\end{lemma}
The proof is given in Appendix \ref{appendix:proof:prop:p223lb}.

\textbf{Observations regarding $\pi(q)$:}
We make some observations about the nature of the stationary distribution $\pi(q)$ in the regime $\Re$, for \INTMC-1, \INTMC-2-2, and \INTMC-2-3.
These observations are significant since these are the cases for which $Q^*(c_{c}) \uparrow \infty$ and the observations can also be used in guiding the derivation of asymptotic lower bounds for discrete time models in \cite[Chapters 4 and 5]{vineeth_thesis}.
\begin{figure}
  \centering
  \includegraphics[width=\jImageWidth,height=\jImageHeight]{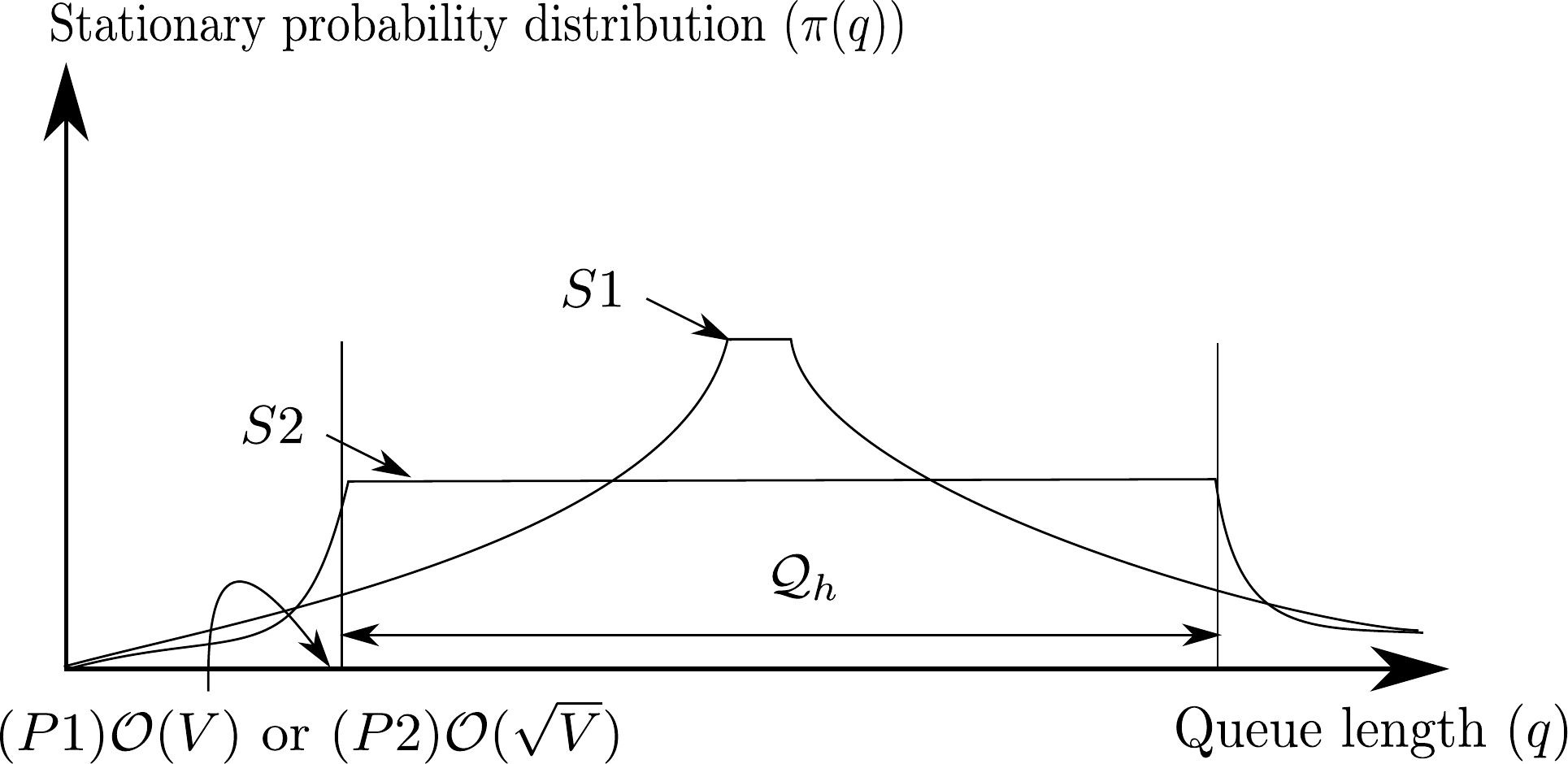}
  \caption{Possibilities for the behaviour of the stationary distribution $\pi(q)$ in the regime $\Re$}
  \label{observations:fig:intuition}
\end{figure}
For the purposes of finding asymptotic lower bounds on the queue length, we note that the elements of $R_{\gamma}$ can be relabelled as starting from zero.
In the following, the state zero corresponds to smallest queue length in $R_{\gamma}$.
We note that in the asymptotic regime $\Re$, for \INTMC-1, \INTMC-2-2, and \INTMC-2-3, $\pi_{\mu}(0) = Pr\brac{\mu(Q) = 0} \downarrow 0$.
Then, we have that $\pi(0) \downarrow 0$.
We have that $\pi(q) \downarrow 0$ for any finite $q \in \Rg$.
Hence, the average queue length should increase.

The difference in the rate at which the average queue length increases for the three cases is due to the difference in the \emph{shape} of $\pi(q)$ in the regime $\Re$ with $V \Deq c_{c} - c(\lambda) \downarrow 0$.
We note that for any admissible policy $\gamma$, the stationary distribution $\pi(q)$ has the behaviour shown in (S1) Figure \ref{observations:fig:intuition}, since $\lambda(q)$ is non-increasing and $\mu(q)$ is non-decreasing as a function of $q$.
For \INTMC-2-2, let $\hpq = \brac{q : \mu(q) \in [a_{\lambda} - \epsilon_{V}, b_{\lambda} + \epsilon_{V}]}$ and for \INTMC-1 and \INTMC-2-3, let $\hpq = \brac{q : \mu(q) \in [\lambda - \epsilon_{V}, \lambda + \epsilon_{V}]}$, where $\epsilon_{V}$ is $\omega(V)$.
Let $q_{1} = \min \hpq$.
For \INTMC-2-2, in the regime $\Re$, $\pi(q)$ retains the shape (S1) since $a_{\lambda} \leq \mu(q) \leq b_{\lambda}, \forall q \in \hpq$.
Then, we have that $\pi(q_{1} - 1) = \mathcal{O}(V)$.
The stationary distribution for $q \in \hpq$ increases geometrically from $q_{1} - 1$ with a geometric factor of at most $\frac{\lambda}{a_{\lambda} - \epsilon_{V}}$.
We note that for \INTMC-1 and \INTMC-2-3, $\mu(q) \rightarrow \lambda$ for $q \in \hpq$ as $V \downarrow 0$.
Then the shape of $\pi(q)$ tends towards (S2) as shown in Figure \ref{observations:fig:intuition}.
For \INTMC-2-3, $\pi(q_{1} - 1)$ is $\mathcal{O}({V})$ (as in (P1) in Figure \ref{observations:fig:intuition}) while for \INTMC-1, $\pi(q_{1} - 1)$ is $\mathcal{O}(\sqrt{V})$ (as in (P2)).
Then, $\pi(q)$ for $q \in \hpq$ tends towards a constant which is $\mathcal{O}(V)$ for \INTMC-2-3 and $\mathcal{O}(\sqrt{V})$ for \INTMC-1.
Therefore, the average queue length scales as $\Omega\nfrac{1}{V}$ and $\Omega\nfrac{1}{\sqrt{V}}$ for \INTMC-2-3 and \INTMC-1 respectively.

\subsection{Asymptotic behaviour of the tradeoff curve}
\label{sec:intmc_ub}
In this section, we present asymptotic upper bounds for the cases \INTMC-1, \INTMC-2-1, \INTMC-2-2, and \INTMC-2-3.
We use these upper bounds along with the asymptotic lower bounds derived in the previous section to obtain a complete asymptotic order characterization of $Q^*(c_{c})$ for \INTMC-2-2 and \INTMC-2-3.
Since in this paper we limit our scope to asymptotic lower bounds, we only present proof outlines for these asymptotic upper bounds.
The complete proofs can be found in \cite[Chapter 3]{vineeth_thesis}.

\begin{lemma}
  For \INTMC-1, there exists a sequence of admissible policies $\gamma_{k}$ with a sequence $V_{k} \downarrow 0$, such that $\overline{Q}(\gamma_{k}) = \mathcal{O}\brap{\frac{1}{\sqrt{V_{k}}}\log\nfrac{1}{V_{k}}}$ and $\overline{C}(\gamma_{k}) - c(\lambda) = V_{k}$.
  \label{prop:p21ub}
\end{lemma}
We note that the asymptotic upper bound above for the sequence $\gamma_{k}$ does not match the asymptotic lower bound $\Omega\nfrac{1}{\sqrt{V_{k}}}$, which was derived in Lemma \ref{prop:p21lb}.
We present an outline of the proof in Appendix \ref{appendix:outline:prop:p21ub}.

\begin{lemma}
  For \INTMC-2-1, there exists a sequence of admissible policies $\gamma_{k}$, with a sequence of $V_{k} \downarrow 0$, such that $\frac{\lambda}{b_{\lambda} - \lambda} - \overline{Q}(\gamma_{k}) = \Theta\brap{V_{k} \log \nfrac{1}{V_{k}}}$ and $\overline{C}(\gamma_{k}) - c(\lambda) = V_{k}$.
  \label{prop:p221ub}
\end{lemma}
We note that the above asymptotic upper bound does not match the asymptotic lower bound in Lemma \ref{prop:p221lb}.
However, if $\XM$ were a discrete set, as in the case of FINITE-$\mu$CHOICE in \cite[Chapter 2]{vineeth_thesis}, then the above asymptotic upper bound matches with the asymptotic lower bound and a complete order characterization can be obtained (see \cite[Proposition 2.3.9]{vineeth_thesis}).
An outline of the proof is given in Appendix \ref{appendix:outline:prop:p221ub}.

\begin{lemma}
  For \INTMC-2-2, there exists a sequence of admissible policies $\gamma_{k}$, with a sequence $V_{k} \downarrow 0$, such that $\overline{Q}(\gamma_{k}) = \mathcal{O}\brap{\log\nfrac{1}{V_{k}}}$ and $\overline{C}(\gamma_{k}) - c(\lambda) = V_{k}$.
  \label{prop:p222ub}
\end{lemma}
An outline of the proof is given in Appendix \ref{appendix:outline:prop:p222ub}.

Using the asymptotic lower bound on $\overline{Q}(\gamma_{k})$ from Lemma \ref{prop:p222lb}, and the asymptotic upper bound above, we obtain the following result.
\begin{proposition}
  For \INTMC-2-2, we have that the optimal tradeoff curve $Q^*(c_{c,k})$ is $\Theta\left(\log\nfrac{1}{c_{c,k} - c(\lambda)}\right)$, for a sequence $c_{c,k} = \overline{C}(\gamma_{k})$, where $\gamma_{k}$ is the sequence of policies in Lemma \ref{prop:p222ub}.
  \label{prop:p222}
\end{proposition}
The proof is given in Appendix \ref{appendix:proof:prop:p222}.

\begin{lemma}
  For \INTMC-2-3, there exists a sequence of admissible policies $\gamma_{k}$, with a sequence $V_{k} \downarrow 0$, such that $\overline{Q}(\gamma_{k}) = \mathcal{O}{\nfrac{1}{V_{k}}}$ and $\overline{C}(\gamma) - c(\lambda) = V_{k}$.
  \label{lemma:p223ub}
\end{lemma}
An outline of the proof is given in Appendix \ref{appendix:outline:p223ub}.

Using the asymptotic lower bound on $\overline{Q}(\gamma_{k})$ from Lemma \ref{prop:p223lb}, the asymptotic upper bound above, and proceeding as in the proof of Proposition \ref{prop:p222} we obtain the following result.
\begin{proposition}
  For \INTMC-2-3, we have that the optimal tradeoff curve $Q^*(c_{c,k})$ is $\Theta\nfrac{1}{c_{c,k} - c(\lambda)}$, for a sequence $c_{c,k} = \overline{C}(\gamma_{k})$, where $\gamma_{k}$ is the sequence of policies in Lemma \ref{lemma:p223ub}.
  \label{prop:p223}
\end{proposition}

For \INTMC\, we note that the arrival rate is $\lambda$.
Hence the minimum average delay as a function of the constraint $c_{c}$ is $\frac{Q^*(c_{c})}{\lambda}$ (from Little's law).
Thus, the asymptotic behaviour of the minimum average delay is the same as that of $Q^*(c_{c})$.

\section{Analysis of \INTLC}
\label{sec:intlc_analysis}
In this section, we obtain asymptotic bounds for \INTLC, which are obtained using techniques similar to those presented above for \INTMC.
We note that for \INTLC\, $\mu(q) = \mu, \forall q \in \brac{1,2,\dots}$ and $\lambda(q) \in [0, r_{a,max}]$.
We note that $\overline{C}(\gamma)$ for $\gamma \in \Gamma_{a}$ is $(1 - \pi(0))c(\mu)$, which depends on the policy, unlike \INTMC\, where the choice of $\lambda$ fixed $\overline{U}(\gamma)$ to be $u(\lambda)$.
For \INTLC, we restrict to admissible $\gamma$ such that $\mu(q) = \mu, \forall q > 0$, where $\mu$ is such that $c(\mu) = c_{c}$, so that $\overline{C}(\gamma) \leq c_{c}$.
The tradeoff problem \INTLC\, is
\begin{eqnarray}
  \text{ minimize }_{\gamma \in \Gamma_{a}} & \overline{Q}(\gamma) \nonumber \\
  \text{ and } & \overline{U}(\gamma) \geq u_{c}.
  \label{chap4:eq:sysmodel:alternate_probstat}
\end{eqnarray}
The optimal value of the above problem is denoted as $Q^*(u_{c})$.
We also note from Lemma \ref{lemma:feasible_solns}, that the maximum value of $\Ug$ over all admissible $\gamma$ is $u(\mu)$.
We obtain an asymptotic characterization of $Q^*(u_{c})$ in the asymptotic regime $\Re$, where $u_{c} \uparrow u(\mu)$.
We note that for $\gamma \in \Gamma_{a}$, $\Ug \leq u(\Exp \lambda(Q)) \leq u(\mu)$.

We assume that $\mu$ is such that $r_{a,max} > \mu$.
If $r_{a,max} = \mu$, then we note that only the non-admissible policy $\gamma$, with $\lambda(q) = r_{a,max}, \forall q$, can achieve $u(\mu)$.
For the asymptotic analysis of \INTLC, we define a line $l(\lambda)$ whose definition is similar to that of $l(\mu)$ for \INTMC.
For \INTLC-1, $l(\lambda)$ is defined as the tangent to $u(\lambda)$ at $\lambda = \mu$.
For \INTLC-2-1, $l(\lambda)$ is defined as the line through $(a_{\mu}, u(a_{\mu}))$ and $(b_{\mu}, u(b_{\mu}))$, while for \INTLC-2-2, $l(\lambda)$ is any line through $(a_{\mu} a_{i}, u(a_{\mu}))$ with slope $m$, such that $ \frac{u(a_{i + 1}) - u(a_{i})}{a_{i + 1} - a_{i}} < m < \frac{u(a_{i}) - u(a_{i - 1})}{a_{i} - a_{i - 1}}$.
We note that $l(\lambda) \geq u(\lambda)$ and $\Exp l(\lambda(Q)) = l(\Exp\lambda(Q))$.
We also note that the function $l(\lambda) - u(\lambda)$ is a convex function.

\begin{lemma}
  For \INTLC, for any sequence of admissible policies such that $u(\mu) - \Ugk = V_{k} \downarrow 0$, we have that $\pi(0) = \mathcal{O}(V_{k})$.
  Therefore, as $u_{c} \uparrow u(\mu)$, $\pi(0) \downarrow 0$, for any sequence of feasible policies for \eqref{chap4:eq:sysmodel:alternate_probstat}.
  \label{lemma:intlc_pi0ub}
\end{lemma}
The proof is given in Appendix \ref{appendix:proof:lemma:intlc_pi0ub}.
Thus, intuitively for problem \eqref{chap4:eq:sysmodel:alternate_probstat} we do not have a case where $Q^*(u_{c})$ increases only up to a finite value as $u_{c} \uparrow u(\mu)$ (unlike \INTMC-2-1).

As for \INTMC, we consider the following cases for \INTLC:
\begin{description}
\item[\INTLC-1:] $u(\lambda)$ is strictly concave for $\lambda \in [0,r_{a,max}]$. 
\item[\INTLC-2:] $u(\lambda)$ is piecewise linear and concave, i.e., (a) there exists a minimal partition of $[0,r_{a,max}]$ into intervals $\{[a_{i},b_{i}], i \in \{1,\dots,P\}\}$ with $a_{1} = 0$, $b_{P} = r_{max}$, and $b_{i} = a_{i + 1}$ and (b) there are linear functions $f_{i}$ such that $\forall \mu \in [a_{i},b_{i}], f_{i}(\lambda) = u(\lambda)$. This is further subdivided into two cases:
  \begin{description}
  \item[1.] $a_{\mu} \Deq a_{i} < \mu < b_{\mu} \Deq b_{i}$, for some $i \geq 1$.
  \item[2.] $a_{\mu} \Deq a_{i} = \mu$, for some $i > 1$.
  \end{description}
\end{description}
We first present asymptotic lower bounds on $Q^*(u_{c})$ in the regime $u_{c} \uparrow u(\mu)$.
Then for \INTLC-1 as well as for \INTLC-2, we show that there exists a sequence of admissible policies $\gamma_{k}$ such that $\Ugk \uparrow u(\mu)$ in Lemma \ref{lemma:intlc_ub}.
Asymptotic lower bounds for \INTLC\, are obtained along similar lines as for \INTMC.
For all cases, we first obtain upper bounds on the stationary probability of certain arrival rates (rather than service rates), which go to zero as $u_{c} \uparrow u(\mu)$.
Then as before, these upper bounds on the stationary probability of certain arrival rates lead to constraints on the stationary probability of all queue lengths.
Since the stationary probability of the queue length determines the average queue length, the constraints determine the behaviour of average queue length as $u_{c} \uparrow u(\mu)$.

For any policy, the set of arrival rates $\brac{\lambda(q) : q \in \sZ}$ is countable and is denoted as $(\lambda_{0}, \lambda_{1}, \dots)$, with $\lambda_{k} < \lambda_{k + 1}$.
For an admissible policy, let $\pi_{\lambda}(k)$ denote the stationary probability of using an arrival rate $\lambda_{k}$, i.e., $\pi_{\lambda}(k) = Pr\brac{\lambda(Q) = \lambda_{k}} = \sum_{\brac{q : \lambda(q) = \lambda_{k}}} \pi(q)$.
We make the following assumption, which is similar to (C2):
\begin{description}
\item[U2:] For \INTLC-1, the second derivative of $u(\lambda)$ at $\lambda = \mu$ is non-zero.
\end{description}
Since the techniques used in the analysis of \INTMC\, and \INTLC\, are similar, we expect that asymptotic upper bounds and lower bounds on any sequence of order-optimal policies can be obtained for \INTLC\, as for \INTMC, with the roles of $\mu(q)$ and $\lambda(q)$ interchanged.
\subsection{Asymptotic lower bounds}
We recall the convention in relabelling the states in $\Rg$ as in Remark \ref{remark:recurrentclass}.
We first obtain bounds on $\pi(q)$ as a function of $u_{c}$ in the following lemma.
\begin{lemma}
  Let $\pi$ be the stationary probability distribution of $Q(t)$ and let $Q \sim \pi$. Let $V \Deq u(\mu) - u_{c}$. Let $S \subseteq [0, r_{a,max}]$ and $Q_{S} = \brac{q : \lambda(q) \in S}$. Then for $\epsilon_{V} > 0$, 
  \begin{enumerate}
    \item for \INTLC-1 with $S = [0, \mu - \epsilon_{V}) \bigcup (\mu + \epsilon_{V}, r_{a,max}]$ we have that $Pr\brac{\lambda(Q) \in S} \leq \frac{V}{a_{1} \epsilon_{V}^{2}}$, where $a_{1} > 0$,
    \item for \INTLC-2-1 with $S = [0, a_{\mu} - \epsilon_{V}) \bigcup (b_{\mu} + \epsilon_{V}, r_{max}]$ we have that $Pr\brac{\lambda(Q) \in S} \leq \frac{V}{m_{a} \epsilon_{V}}$,
    \item for \INTLC-2-2 with $S = [0, \mu - \epsilon_{V}) \bigcup (\mu + \epsilon_{V}, r_{max}]$ we have that $Pr\brac{\lambda(Q) \in S} \leq \frac{V}{m_{a} \epsilon_{V}}$,
  \end{enumerate}
  where $m_{a} > 0$.
  Since $Pr\brac{Q \in Q_{S}} = Pr\brac{\mu(Q) \in S}$, the same bounds hold for $Pr\brac{Q \in Q_{S}}$.
  We note that similar bounds hold for any subset of $S$.
  \label{lemma:statprobbound:lc}
\end{lemma}
The proof is given in Appendix \ref{appendix:proof:lemma:statprobbound:lc}.

\begin{lemma}
  For \INTLC-1, for any sequence of admissible policies $\gamma_{k}$ such that $u(\mu) - \Ugk = V_{k} \downarrow 0$, we have that $\Qgk = \Omega{\nfrac{1}{\sqrt{V_{k}}}}$. Therefore, $Q^*(u_{c}) = \Omega{\nfrac{1}{\sqrt{u(\mu) - u_{c}}}}$.
  \label{lemma:intlc-1:lb}
\end{lemma}
The proof is given in Appendix \ref{appendix:proof:lemma:intlc-1:lb}.

\begin{lemma}
  For \INTLC-2-1, then for any sequence of admissible policies $\gamma_{k}$ such that $u(\mu) - \Ugk = V_{k} \downarrow 0$, we have that $\Qgk = \Omega\brap{\log\nfrac{1}{{V_{k}}}}$. Therefore $Q^*(u_{c}) = \Omega\brap{\log\nfrac{1}{{u(\mu) - u_{c}}}}$.
  \label{lemma:intlc-2-1:lb}
\end{lemma}
The proof is given in Appendix \ref{appendix:proof:lemma:intlc-2-1:lb}.

\begin{remark}
  We note that the above asymptotic lower bound holds even in the case where $a_{\mu} = 0$.
  For \INTMC-2, we note that the case with $a_{\lambda} = 0$ corresponds to the case \INTMC-2-1, for which $Q^*(c_{c})$ only increased to a finite value.
\end{remark}

\begin{remark}
  We note that in many cases, for queueing models with a single queue, the utility constraint is on the average throughput.
  Then we have that $u(\lambda)$ is a line segment, with $a_{\mu} = 0$ and $b_{\mu} = r_{a,max}$.
  We note that the asymptotic $\Omega\brap{\log\nfrac{1}{V}}$ lower bound holds for $Q^*(u_{c})$, from the above result.
\end{remark}

\begin{lemma}
  For \INTLC-2-2, for any sequence of admissible policies $\gamma_{k}$ such that $u(\mu) - \Ugk = V_{k} \downarrow 0$, we have that $\Qgk = \Omega{\nfrac{1}{{V_{k}}}}$.
  \label{lemma:intlc-2-2:lb}
\end{lemma}
The proof is given in Appendix \ref{appendix:proof:lemma:intlc-2-2:lb}.

\textbf{Observations regarding $\pi(q)$:}
For \INTLC, from Lemma \ref{lemma:intlc_pi0ub}, we have that $\pi(0) \downarrow 0$ in the asymptotic regime $\Re$ where $u_{c} \uparrow u(\mu)$.
Then, as in the case of \INTMC, we have that $\pi(q)$ for any finite $q \in \Rg$, decreases to $0$ in the regime $\Re$.
Therefore, the average queue length has to increase to infinity.
We note that the observations for the asymptotic behaviour of \INTLC\, and \INTMC\, are similar except that the roles of $\mu(q)$ and $\lambda(q)$ are interchanged.
Again the different asymptotic behaviours of $Q^*(u_{c})$ can be attributed to the different behaviours of $\pi(q)$ as shown in Figure \ref{observations:fig:intuition}.
\subsection{Asymptotic upper bound}
We note that as for \INTMC, using policies with similar structure as in Lemmas \ref{prop:p21ub}, \ref{prop:p222ub}, and \ref{lemma:p223ub} it is possible to obtain a $\mathcal{O}\brap{\frac{1}{\sqrt{V}}\log\nfrac{1}{V}}$ asymptotic upper bound for \INTLC-1 and tight asymptotic upper bounds for \INTLC-2-1 and \INTLC-2-2.
Here, we present a single asymptotic upper bound for \INTLC-1, \INTLC-2-1, and \INTLC-2-2, which shows that there exists a sequence of admissible policies $\gamma_{k}$ such that $\Ugk \uparrow u(\mu)$.
\begin{lemma}
  There exists a sequence of admissible policies $\gamma_{k}$ such that $\Qgk = \mathcal{O}\nfrac{1}{V_{k}}$ and $u(\mu) - \Ugk = V_{k}$.
  \label{lemma:intlc_ub}
\end{lemma}
Since in this paper, we limit our scope to asymptotic lower bounds we skip the proof.
The proof can be found in \cite[Lemma 3.2.28]{vineeth_thesis}.

For \INTLC\, we note that the average arrival rate $\Exp \lambda(Q)$ for any feasible admissible policy satisfies
\begin{eqnarray*}
 u^{-1}(u_{c}) \leq \Exp \lambda(Q) \leq \mu.
\end{eqnarray*}
Therefore, the minimum average delay as a function of the constraint $u_{c}$, $D^*(u_{c})$, satisfies
\begin{eqnarray*}
 \frac{Q^*(u_{c})}{\mu} \leq D^*(u_{c}) \leq \frac{Q^*(u_{c})}{u^{-1}(u_{c})}.
\end{eqnarray*}
In the asymptotic regime $\Re$, for a sequence $u_{c,k} \uparrow u(\mu)$, we note that a uniform upper bound on $D^*(u_{c,k})$ is $\frac{Q^*(u_{c})}{u^{-1}(u_{c,1})}$.
Thus, the asymptotic behaviour of $D^*(u_{c})$ is the same as that of $Q^*(u_{c})$.

\section{Analysis of \INTLMC}
\label{sec:problemp3}

We recall that for \INTLMC, $\lambda(q) \in [r_{a,min},r_{a,max}]$ and $\mu(q) \in [0,r_{max}]$, $\forall q \in \mathbb{Z}_{+}$, with $r_{a,min} < r_{max}$.
The tradeoff problem for \INTLMC\, is:
\begin{eqnarray}
  \text{ minimize }_{\gamma \in \Gamma_{a}} & \overline{Q}(\gamma) \nonumber \\
  \text{ such that } & \overline{C}(\gamma) \leq c_{c}, \nonumber \\
  \text{ and } & \overline{U}(\gamma) \geq u_{c}.
  \label{chap4:eq:intlmc_problem_statement}
\end{eqnarray}
The optimal value of the above problem is denoted as $Q^*(c_{c}, u_{c})$.
Although it is possible to consider various forms of the function $c(\mu)$ as in the case of \INTMC\, and \INTLC\, here we present a complete analysis for the case where  $c(\mu)$ is a strictly convex function of $\mu \in [0, r_{max}]$ (with assumption C2) and $u(\lambda)$ is either a strictly concave (with assumption U2) or a piecewise linear function of $\lambda \in [r_{a,min}, r_{a,max}]$.
We then comment on the asymptotic bounds for other forms of $c(\mu)$ in the following discussion.
\begin{remark}
  In this analysis, we assume that $u_{c} \leq u(r_{a,max})$.
  If $u_{c} > u(r_{a,max})$, then there does not exist any feasible policies for \eqref{chap4:eq:intlmc_problem_statement}.
  We note that if $u_{c} = u(r_{a,max})$, then policies which satisfy this utility constraint need to have $\lz = r_{a,max}, \forall q$, and if $r_{a,max} < r_{max}$ the problem is the same as that considered in \INTMC-1.
\end{remark}
We recall the convention in relabelling the states in $\Rg$ as in Remark \ref{remark:recurrentclass}.
We note that for any feasible policy for \INTLMC, $\Cg \geq c(u^{-1}(u_{c}))$ from Lemma \ref{lemma:feasible_solns}.
We consider \INTLMC\, in the asymptotic regime $\Re$ where $c_{c,k}$ approaches $c(u^{-1}(u_{c,k}))$.
We first consider the special case where $u_{c,k}$ is fixed and comment on the other cases in Section \ref{chap4:sec:sim_probs_INTMC}.

\subsection{Asymptotic lower bound}
In this section we present an asymptotic lower bound on $\overline{Q}(\gamma_{k})$ for any sequence of admissible policies $\gamma_{k}$ for which $\overline{U}(\gamma_{k}) \geq u_{c}$ and $\overline{C}(\gamma_{k}) \downarrow c(u^{-1}(u_{c}))$.
Subsequently, in Lemma \ref{prop:p3ub} we show that there exists a sequence of admissible policies $\gamma_{k}$ for which $\overline{C}(\gamma_{k})$ approaches $c(u^{-1}(u_{c}))$ arbitrarily closely.
\begin{lemma}
  For \INTLMC, for any sequence of admissible policies $\gamma_{k}$ such that $\overline{C}(\gamma_{k}) - c(u^{-1}(u_{c})) = V_{k} \downarrow 0$ and $\overline{U}(\gamma_{k}) \geq u_{c}$, we have that
  \begin{eqnarray*}
    \overline{Q}(\gamma_{k}) = \Omega\brap{\log\nfrac{1}{V_{k}}}.
  \end{eqnarray*}
  \label{prop:p3lb}
\end{lemma}
The proof is given in Appendix \ref{appendix:proof:prop:p3lb}.
We note that in the proof, no use is made of the assumption that the sequence of policies satisfies the constraint $\Expp u(\lambda(Q)) \geq u_{c}$.
The difficulty with \INTLMC\, is in actually constructing a sequence of policies which achieves the above asymptotic lower bound.

\textbf{Observations regarding $\pi(q)$:}
We again have that $\pi(0) \downarrow 0$ in the asymptotic regime $\Re$ where $V \Deq c_{c} - c(\newl) \downarrow 0$.
Then, as in the case of \INTMC\, or \INTLC\, we have that $\pi(q)$ for any finite $q \in \Rg$, decreases to $0$ in the regime $\Re$.
Therefore, the average queue length has to increase to infinity.
We note that as in the case of \INTMC-1 or \INTMC-2-3, a set of queue lengths $\hpq$ occurring with high probability can be defined as $\hpq = \brac{q : \mu(q) \in [\newl - \epsilon_{V}, \newl + \epsilon_{V}]}$, where $\epsilon_{V}$ is $\omega(V)$.
We note that as $V \downarrow 0$, $\mu(q)$ for $q \in \hpq$ tends to $\newl$.
However, since $\lambda(q)$ can be controlled, unlike \INTMC-1 or \INTMC-2-3, for \INTLMC\, the behaviour of $\pi(q)$ is as in (S1) in Figure \ref{observations:fig:intuition} rather than (S2).
Then, the average queue length scales only as $\Omega\brap{\log\nfrac{1}{V}}$.

\subsection{Asymptotic behaviour of the tradeoff curve}
In this section, we construct a sequence of admissible policies $\gamma_{k}$ which achieves $c(\newl)$ arbitrarily closely with $\overline{Q}(\gamma_{k})$ scaling as in Lemma \ref{prop:p3lb}, showing that the scaling is optimal.
However, we are able to obtain an asymptotic upper bound only for the case where $u(\lambda)$ is strictly concave or linear (and not piecewise linear).
\begin{lemma}
  For \INTLMC, with $u(\lambda)$ strictly concave or linear, there exists a sequence of admissible policies $\gamma_{k}$ with a corresponding sequence $V_{k} \downarrow 0$ such that
  \begin{eqnarray*}
    \overline{Q}(\gamma_{k}) & = & \mathcal{O}\brap{\log\nfrac{1}{V_{k}}},\\
    \overline{C}(\gamma_{k}) - c(\newl) & = & V_{k}, \\
    \overline{U}(\gamma_{k}) & \geq & u_{c},
  \end{eqnarray*}
  for any $0 < u_{c} < u(r_{a, max})$.
  \label{prop:p3ub}
\end{lemma}
The proof is given in Appendix \ref{appendix:proof:prop:p3ub}.
The construction of the sequence of admissible policies $\gamma_{k}$ is motivated by the following intuition, that we have obtained from the lower bound in Lemma \ref{prop:p3lb}.
The sequence of policies should be such that as $V_{k} \downarrow 0$, the service rate used, at a queue length in $\hpq$, should be close to $\newl$.
But the arrival rate $\lambda(q)$ should not exactly equal $\newl$, for all queue lengths $q \in \hpq$.
Then it should be possible to have a stationary distribution which is geometrically growing and then decaying, leading to the required $\log\nfrac{1}{V_{k}}$ scaling of $\overline{Q}(\gamma_{k})$.

\begin{remark}
  We note that the above proof also applies if $u(\lambda)$ is piecewise linear and $(\newl, u_{c})$ lies on a linear segment of the piecewise linear function $u(\lambda)$.
  However, the proof does not apply if $u(\lambda)$ is piecewise linear and $u_{c}$ is such that the slope of $u(\lambda)$ changes at $(\newl, u_{c})$.
\end{remark}

Using the asymptotic lower bound from Lemma \ref{prop:p3lb}, the asymptotic upper bound above, we arrive at the following result.

\begin{proposition}
  For \INTLMC, for strictly concave or linear $u(\lambda)$, we have that the optimal tradeoff curve $Q^*(c_{c,k},u_{c}) = \Theta\brap{\log\nfrac{1}{c_{c,k} - c(u^{-1}(u_{c}))}}$, for the sequence $c_{c,k} = \Cgk$, for the sequence of policies $\gamma_{k}$ in Lemma \ref{prop:p3ub}.
  \label{prop:p3}
\end{proposition}

\begin{remark}
  We note that for \INTLMC\, we have considered the case where $c(\mu)$ is strictly convex and $u(\lambda)$ is either strictly concave or linear (also piecewise linear for the asymptotic lower bound in Lemma \ref{prop:p3lb}).
  Although we have not presented the analysis for other forms of $c(\mu)$, such as when $c(\mu)$ is piecewise linear, here we outline how the methods presented above can be used in obtaining asymptotic lower bounds in these cases, in the asymptotic regime where $c_{c,k} \downarrow c(\newl)$.
  Suppose $c(\mu)$ is piecewise linear.
  We note that as in \INTMC-2, we can define service rates $a_{\lambda}$ and $b_{\lambda}$ with respect to $u^{-1}(u_{c})$ rather than $\lambda$.
  Then the asymptotic behaviour of $Q^*(c_{c},u_{c})$ depends upon whether (i) $a_{\lambda} < u^{-1}(u_{c}) < b_{\lambda}$ and $a_{\lambda} = 0$ or (ii) otherwise.
  For case (i), we note that $Q^*(c_{c}, u_{c})$ only increases to a finite value, since we can fix $\lambda(q) = u^{-1}(u_{c})$ and apply the analysis of \INTMC-2-1.
  However, we do not have an asymptotic lower bound in this case.
  For case (ii), we can proceed as in the proof of Lemma \ref{prop:p3lb}, except that $\mu^* \Deq \mu_{l} - \epsilon$, where $\epsilon > 0$, to obtain that $Q^*(c_{c}, u_{c})$ is $\Omega\brap{\log\nfrac{1}{V}}$.
\end{remark}

For \INTLMC\, we note that the average arrival rate $\Exp \lambda(Q)$ for any feasible admissible policy satisfies
\begin{eqnarray*}
 u^{-1}(u_{c}) \leq \Exp \lambda(Q) \leq c^{-1}(c_{c}).
\end{eqnarray*}
Therefore, the minimum average delay as a function of the constraints $c_{c}$ and $u_{c}$, $D^*(c_{c}, u_{c})$, satisfies
\begin{eqnarray*}
 \frac{Q^*(c_{c}, u_{c})}{c^{-1}(c_{c})} \leq D^*(u_{c}) \leq \frac{Q^*(c_{c}, u_{c})}{u^{-1}(u_{c})}.
\end{eqnarray*}
We consider the asymptotic regime $\Re$, with $c_{c,k}$ and $u_{c,k}$ such that $c_{c,k} \downarrow c(u^{-1}(u_{c,k}))$.
We assume that there exists $c_{c,1} = \max_{k} c_{c,k}$ and $u_{c,1} = \min_{k} u_{c,k}$.
Then we have that 
\begin{eqnarray*}
 \frac{Q^*(c_{c,k}, u_{c,k})}{c^{-1}(c_{c,1})} \leq D^*(u_{c}) \leq \frac{Q^*(c_{c,k}, u_{c,k})}{u^{-1}(u_{c,1})}.
\end{eqnarray*}
Thus, the asymptotic behaviour of $D^*(c_{c},u_{c})$ is the same as that of $Q^*(c_{c},u_{c})$.

\subsection{Other asymptotic regimes for \INTLMC}
\label{chap4:sec:sim_probs_INTMC}
In our discussion of \INTLMC, the utility constraint $u_{c}$ was kept fixed while $c_{c,k} \downarrow c(u^{-1}(u_{c}))$.
A similar problem (SP1) is one in which $c_{c}$ is fixed and $u_{c,k} \uparrow u(c^{-1}(c_{c}))$.
Another problem scenario (SP2) is one in which both $c_{c,k}$ and $u_{c,k}$ vary such that (a) $c_{c,k} - c(u^{-1}(u_{c,k})) \downarrow 0$ or (b) $u(c^{-1}(c_{c,k})) - u_{c,k} \downarrow 0$.
We note that SP2(b) encompasses SP1 since the sequence $c_{c,k}$ can be chosen such that $c_{c,k} = c_{c}, \forall k \in \sZ$.
In Appendix \ref{appendix:proof:sp2b_sp2a} we show that the asymptotic regime for SP2(b) is equivalent to that for SP2(a), i.e., $c_{c,k} - c\brap{u^{-1}(u_{c,k})} \downarrow 0$.

We have the following result, under the stronger assumption that $u(\lambda)$ is $m$-strongly concave \cite[Section 9.1.2]{boyd}, with $m > 0$.
The proof is similar to that of Lemma \ref{prop:p3lb}.
\begin{lemma}
  For \INTLMC, for any sequence of admissible policies $\gamma_{k}$ and a sequence $u_{c,k} > 0$ such that $\overline{C}(\gamma_{k}) - c(u^{-1}(u_{c,k})) = V_{k} \downarrow 0$ and $\overline{U}(\gamma_{k}) \geq u_{c,k}$, we have that
  \begin{eqnarray*}
    \overline{Q}(\gamma_{k}) = \Omega\brap{\log\nfrac{1}{V_{k}}}.
  \end{eqnarray*}  
  \label{lemma:intlmc-altregime}
\end{lemma}
The proof is given in Appendix \ref{appendix:proof:lemma:intlmc-altregime}.
We note that an asymptotic upper bound can be obtained by evaluating $\Qgk, \Cgk$, and $\Ugk$ for a sequence of policies $\gamma_{k}$ as in Lemma \ref{prop:p3ub}, but with $u_{c}$ now being the sequence $u_{c,k}$.
Then we have the following result
\begin{proposition}
  For \INTLMC, for strongly concave or linear $u(\lambda)$, we have that the optimal tradeoff curve $Q^*(c_{c,k},u_{c,k}) = \Theta\brap{\log\nfrac{1}{c_{c,k} - c(u^{-1}(u_{c,k}))}}$, for the sequence $c_{c,k} = \Cgk$ and $u_{c,k} = \Ugk$, for the sequence of policies $\gamma_{k}$ as above.
\end{proposition}

\section{Discussion}
\subsection{Solutions for \eqref{introduction:eq:mm1_genTradeoff} from TRADEOFF}

\label{sec:discussion}
We note that TRADEOFF is a specific case of \eqref{introduction:eq:mm1_genTradeoff}, obtained by restricting to the set of admissible policies.
As discussed before, $Q^*(c_{c}, u_{c})$ is optimal for \eqref{introduction:eq:mm1_genTradeoff} whenever $(c_{c}, u_{c}) \in \mathcal{O}^{d}_{u}$.
In this section, we discuss how asymptotic lower bounds can be obtained for \eqref{introduction:eq:mm1_genTradeoff} for other values of the constraints $c_{c}$ and $u_{c}$.

We note that the optimal solution of \eqref{introduction:eq:mm1_genTradeoff} is lower bounded by the optimal value of its Lagrange dual
\begin{eqnarray*}
    \max_{\beta_{1} \geq 0, \beta_{2} \geq 0} \bras{\min_{\gamma \in \Gamma} \brac{\Qg + \beta_{1}\brap{\Cg - c_{c}} - \beta_{2}\brap{\Ug - u_{c}}}}.
\end{eqnarray*}
We note that there exists an admissible policy which is optimal for the inner optimization over the set of policies.
Then using the asymptotic lower bounds which were obtained for TRADEOFF, it is possible to obtain asymptotic lower bounds for the above problem.
For example, for \INTMC-2-2, we have the following result.
\begin{proposition}
Suppose $\gamma_{k}$ is any sequence of policies such that $\Cgk - c(\lambda) = V_{k} \downarrow 0$ and $\Qgk = \mathcal{O}\brap{\log\nfrac{1}{V_{k}}}$. Let $c_{c,k}$ be any sequence such that $c_{c,k} = \Theta\brap{\Cgk - c(\lambda)}$. Then the optimal value of \eqref{introduction:eq:mm1_genTradeoff} is $\Omega\brap{\log\nfrac{1}{c_{c,k} - c(\lambda)}}$.
\label{prop:duallowerbound}
\end{proposition}
The proof is given in Appendix \ref{appendix:proof:prop:duallowerbound}. Similar results are presented in \cite[Proposition 4.3.22]{vineeth_thesis}. 
For any sequence $c_{c,k} \in \mathcal{O}^{d}_{u}$, there exists a sequence of policies which satisfies the requirements for $\gamma_{k}$ in the above proposition.
We note that the above result only requires that a sequence of policies, admissible or otherwise, exists with the stated properties.
Thus, we have asymptotic lower bounds for \eqref{introduction:eq:mm1_genTradeoff} for a set of $c_{c}$ which contains $\mathcal{O}^{d}_{u}$.

\subsection{Tradeoff for \emph{mixture} policies}

We now consider a tradeoff problem, where the set of policies also include policies which are obtained by mixing the \emph{pure} policies in $\Gamma_{a}$.
The set of policies which are obtained by a finite mixture of the policies in $\Gamma_{a}$ is denoted as $\Gamma_{a,M}$.
We note that associated with a $\gamma_{M} \in \Gamma_{a,M}$ we have a set $\Gamma(\gamma_{M}) \subseteq \Gamma_{a}$, which is the set of pure policies which are mixed according to a probability mass function $p_{\gamma}$, for $\gamma \in \Gamma(\gamma_{M})$.
For a $\gamma_{M} \in \Gamma_{a,M}$, $\overline{Q}(\gamma_{M}) = \sum_{\gamma} p_{\gamma} \Qg$ (which is also denoted as $\Exp_{p_{\gamma}} \Qg$).
The average service cost rate and average utility rate are defined similarly for $\gamma \in \Gamma_{a,M}$.
For $\Gamma_{a,M}$ we have the following optimization problem, TRADEOFF-M:
\begin{eqnarray}
  \text{ minimize }_{\gamma \in \Gamma_{a,M}} & \overline{Q}(\gamma) \nonumber \\
  \text{ such that } & \overline{C}(\gamma) \leq c_{c}, \nonumber \\
  \text{ and } & \overline{U}(\gamma) \geq u_{c},
  \label{chap4:eq:sysmodel:probstat}
\end{eqnarray}
where $c_{c}$ and $u_{c}$ are constraints on the average service cost rate and average utility rate respectively.
The optimal value of the above problem is denoted by $Q^*_{M}(c_{c},u_{c})$.
Asymptotic bounds on $Q^*_{M}(c_{c}, u_{c})$ can be obtained from $Q^*(c_{c}, u_{c})$ (e.g. \cite[Proposition 2.3.9]{vineeth_thesis}) and are the same as that of $Q^*(c_{c}, u_{c})$.

\section{Summary and Conclusions}
\label{sec:summary}
In this paper, we have obtained asymptotic lower bounds and presented asymptotic upper bounds for the tradeoff of average queue length, with average service cost rate and average utility rate for a continuous time state dependent M/M/1 queueing model.
The results that we have presented in this paper are summarized in Table \ref{conclusions:table:results}.
The asymptotic lower bounds are obtained by considering the stationary distribution $\pi(q)$ of the queue length in the asymptotic regime $\Re$, with $V = c_{c} - c(u^{-1}(u_{c})) \downarrow 0$.
We presented some observations about the behaviour of $\pi(q)$ in the regime $\Re$.
In the regime $\Re$, the difference in the behaviour of $\pi(q)$ leads to different asymptotic growth rates for the average queue length and thus the average delay.
The general observations regarding the behaviour of $\pi(q)$ that we obtain are significant - the observations are used in obtaining bounds on the stationary probability distribution of queue length for discrete time queueing models.
These bounds have similar behaviour as $\pi(q)$ in a similar asymptotic regime $\Re$.
The bounds on the stationary probability distribution of queue length for discrete time queueing models are then used in deriving similar asymptotic lower bounds.
For example, in \cite[Chapter 5]{vineeth_thesis} we rederive the $\Omega\nfrac{1}{\sqrt{V}}$ Berry-Gallager asymptotic lower bound for admissible policies for a discrete time queueing model, with a strictly convex service cost function.
We also obtain new asymptotic lower bounds or rederive known asymptotic lower bounds for other discrete time queueing models in \cite[Chapters 4 and 5]{vineeth_thesis}.
We also note that our approach using bounds on $\pi(q)$ is also useful in obtaining asymptotic bounds on the structure of order-optimal admissible policies for the state dependent M/M/1 model.
We obtain an asymptotic characterization of the cardinality of the set $Q_{S}$, where $Q_{S} = \brac{q:\mu(q) \in S}$ (e.g. $S = [0, \lambda - V^{\frac{1}{2}}]$ for \INTMC-1) in \cite[Proposition 2.3.16, Lemma 3.2.18]{vineeth_thesis}.
These asymptotic bounds on the cardinality of $Q_{S}$ are useful in obtaining guidance for designing order-optimal admissible policies for both the state dependent M/M/1 model as well as discrete time models as in \cite[Chapters 4 and 5]{vineeth_thesis}.

\begin{table}
  \centering
  \begin{tabular}{|l|l|c|}
    \hline
    \textbf{Control} & \textbf{Service cost and utility functions} & 
    \begin{minipage}{0.3\textwidth}
      \textbf{Results (in the regime $\Re$, for admissible policies)}
    \end{minipage} \\
    \hline
    \begin{minipage}{0.3\textwidth}
      $\mathcal{X}_{\mu} = [0, r_{max}]$, $\mathcal{X}_{\lambda} = \brac{\lambda}$
    \end{minipage} & 
    \begin{minipage}{0.3\textwidth}
      $c(\mu)$ is strictly convex
    \end{minipage} &
    \begin{minipage}{0.3\textwidth}
      \vspace{0.1em}
      Minimum average queue length is $\Omega\nfrac{1}{\sqrt{V}}$ 
    \end{minipage} \\
    &
    \begin{minipage}{0.3\textwidth}
      $c(\mu)$ is piecewise linear
    \end{minipage} &
    \begin{minipage}{0.3\textwidth}
      \vspace{0.2em}
      Depending on $\lambda$, minimum average queue length either increases to a finite value, is $\Theta\brap{\log\nfrac{1}{V}}$, or $\Theta\nfrac{1}{V}$
    \end{minipage} \\
    \hline
    \begin{minipage}{0.3\textwidth}
      $\mathcal{X}_{\mu} = \brac{\mu}$, $\mathcal{X}_{\lambda} = [0, r_{a,max}]$
    \end{minipage} & 
    \begin{minipage}{0.3\textwidth}
      $u(.)$ is strictly concave
    \end{minipage} &
    \begin{minipage}{0.3\textwidth}
      \vspace{0.1em}
      Minimum average queue length is $\Omega\nfrac{1}{\sqrt{V}}$ 
    \end{minipage} \\
    &
    \begin{minipage}{0.3\textwidth}
      $u(.)$ is piecewise linear
    \end{minipage} &
    \begin{minipage}{0.3\textwidth}
      \vspace{0.2em}
      Depending on $\lambda$, minimum average queue length is $\Omega\brap{\log\nfrac{1}{V}}$, or $\Omega\nfrac{1}{V}$
    \end{minipage} \\
    \hline
    \begin{minipage}{0.3\textwidth}
      $\mathcal{X}_{\mu} = [0, r_{max}]$, $\mathcal{X}_{\lambda} = [0, r_{a,max}]$ 
    \end{minipage} &
    \begin{minipage}{0.3\textwidth}
      $c(.)$ is strictly convex, $u(.)$ is strictly concave
    \end{minipage}
    &
    \begin{minipage}{0.3\textwidth}
      Minimum average queue length is $\Theta\brap{\log\nfrac{1}{V}}$
    \end{minipage} \\
    \hline
  \end{tabular}
  \caption{Asymptotic results derived in this paper}
  \label{conclusions:table:results}
\end{table}

\bibliographystyle{plain}
\bibliography{mm1}

\begin{thebibliography}{10}

\bibitem{ata_pcstatic}
B.~Ata.
\newblock Dynamic power control in a wireless static channel subject to a
  quality-of-service constraint.
\newblock {\em Operations Research}, 53(5), 2005.

\bibitem{ata}
B.~Ata, J.~M. Harrison, and L.~A. Shepp.
\newblock Drift rate control of a {Brownian} processing system.
\newblock {\em The Annals of Applied Probability}, 15(2), 2005.

\bibitem{atamm1}
B.~Ata and S.~Shneorson.
\newblock Dynamic control of a {M/M/1} service system with adjustable arrival
  and service rates.
\newblock {\em Management Science}, 52(11), 2006.

\bibitem{berry_thesis}
R.A. Berry.
\newblock {\em Power and delay tradeoffs in fading channels}.
\newblock PhD thesis, Laboratory for Information and Decision Systems,
  Massachusetts Institute of Technology, 2000.

\bibitem{berry}
R.A. Berry and R.G. Gallager.
\newblock Communication over fading channels with delay constraints.
\newblock {\em IEEE Transactions on Information Theory}, 48(5), May 2002.

\bibitem{boyd}
S.~Boyd and L.~Vandenberghe.
\newblock {\em Convex Optimization}.
\newblock Cambridge University Press, 2004.

\bibitem{chaporkar}
P.~Chaporkar, K.~Kar, Xiang Luo, and S.~Sarkar.
\newblock Throughput and fairness guarantees through maximal scheduling in
  wireless networks.
\newblock {\em IEEE Transactions on Information Theory}, 54(2), 2008.

\bibitem{george}
J.~M. George and J.~M. Harrison.
\newblock Dynamic control of a queue with adjustable service rate.
\newblock {\em Operations Research}, 49(5), 2001.

\bibitem{hernandez}
J.~Gonzalez-Hernandez and C.~E. Villarreal.
\newblock Optimal policies for constrained average-cost {Markov} decision
  processes.
\newblock {\em TOP Journal of Spanish Society of Statistics and Operations
  Research}, 19(1), July 2011.

\bibitem{koole}
G.~Koole.
\newblock {\em Monotonicity in Markov reward and decision chains: theory and
  applications}.
\newblock Foundations and Trends in Stochastic Systems: NOW Publishers, 2007.

\bibitem{ma}
D.J. Ma, A.~M. Makowski, and A.~Shwartz.
\newblock Estimation and optimal control for constrained {Markov} chains.
\newblock In {\em Proceedings of the 25th IEEE Conference on Decision and
  Control}, 1986.

\bibitem{neely_mac}
M.J. Neely.
\newblock Optimal energy and delay tradeoffs for multiuser wireless downlinks.
\newblock {\em IEEE Transactions on Information Theory}, 53(9), Sept. 2007.

\bibitem{neely_utility}
M.J. Neely.
\newblock Intelligent packet dropping for optimal energy-delay tradeoffs in
  wireless downlinks.
\newblock {\em IEEE Transactions on Automatic Control}, 54(3), March 2009.

\bibitem{neely}
M.J. Neely.
\newblock {\em Stochastic network optimization with application to
  communication and queueing systems}.
\newblock Morgan and Claypool, 2010.

\bibitem{perkins}
J.R. Perkins and R.~Srikant.
\newblock The role of queue length information in congestion control and
  resource pricing.
\newblock In {\em Proceedings of the 38th IEEE Conference on Decision and
  Control}. IEEE, 1999.

\bibitem{venkithesis}
V.~Ramaiyan.
\newblock {\em Topics in modelling, analysis and optimization of wireless
  networks}.
\newblock PhD thesis, Dept. of Electrical Communication Engineering, Indian
  Institute of Science, 2009.

\bibitem{ross}
Sheldon~M. Ross.
\newblock Average cost semi-markov decision processes.
\newblock {\em Journal of Applied Probability}, 7(3):649--656, 1970.

\bibitem{vineeth_thesis}
Vineeth~B. S.
\newblock {\em On the tradeoff of average delay, average service cost, and
  average utility for single server queues with monotone policies}.
\newblock PhD thesis, Dept. of Electrical Communication Engg., Indian Institute
  of Science, 2013.

\bibitem{vineeth_ncc13_dt}
Vineeth~B. S. and U.~Mukherji.
\newblock Tradeoff of average power and average delay for a point-to-point link
  with fading.
\newblock In {\em Proceedings of the National Conference on Communications
  (NCC), New Delhi, India}, 2013.

\bibitem{vineeth_ncc13_ct}
Vineeth~B. S. and U.~Mukherji.
\newblock Tradeoff of average service cost and average delay for the state
  dependent {M/M/1} queue.
\newblock In {\em Proceedings of the National Conference on Communications
  (NCC), New Delhi, India}, 2013.

\bibitem{sennott}
L.~I. Sennott.
\newblock Constrained average cost {Markov} decision chains.
\newblock {\em Probability in the Engineering and Informational Sciences}, 7,
  1993.

\bibitem{sennott_text}
L.~I. Sennott.
\newblock {\em Stochastic dynamic programming and the control of queues}.
\newblock Wiley- IEEE, 1999.

\bibitem{weber}
S.~Stidham and R.R. Weber.
\newblock Monotonic and insensitive optimal policies for the control of queues
  with undiscounted costs.
\newblock {\em Operations Research}, 37, 1989.

\end{thebibliography}

\begin{appendices}
\section{Proofs}
\subsection{Proof of Lemma \ref{lemma:feasible_solns}}
\label{appendix:proof:lemma:feasible_solns}
We assume that there is an policy $\gamma_{M} \in \Gamma_{a,M}$ which is feasible for TRADEOFF.
For brevity, let us denote $\Exp_{p_{\gamma}} \bras{\Exp_{\pi_{\gamma}}\bras{.}}$ by just $\Exp\bras{.}$ in this proof.
From Jensen's inequality we have that $c(\Exp{\bras{\mu(Q)}}) \leq \Exp\bras{c(\mu(Q))}$ and $\Exp\bras{u(\lambda(Q))} \leq u(\Exp\bras{ \lambda(Q)})$.
Therefore $\Exp \mu(Q) \leq c^{-1}(\Exp c(\mu(Q)))$ and $u^{-1}(\Exp u(\lambda(Q))) \leq \Exp \lambda(Q)$.
As $\Exp_{\pi_{\gamma}} Q < \infty$, $\Exp_{\pi_{\gamma}} \mu(Q) = \Exp_{\pi_{\gamma}} \lambda(Q)$, $\forall \gamma \in \Gamma(\gamma_{M})$.
Therefore for $\gamma$, $u^{-1}(\Exp u(\lambda(Q))) \leq c^{-1}(\Exp c(\mu(Q)))$.
From the non-decreasing properties of $c(.)$ and $u(.)$ we have that $c^{-1}(.)$ and $u^{-1}(.)$ are also non-decreasing.
Hence, if there is any one feasible policy $\gamma$, then $u^{-1}(u_{c}) \leq c^{-1}(c_{c})$. \qeda

\subsection{Proof of Lemma \ref{lemma:statprobbound}}
\label{appendix:proof:lemma:statprobbound}
We first consider \INTMC-1.
We have that
\begin{eqnarray}
  V & = & \sum_{k = 0}^{\infty} \left(c(\mu_{k}) - l(\mu_{k})\right)\pi_{\mu}(k) = \sum_{q = 0}^{\infty} \left( c(\mu(q)) - l(\mu(q))\right) \pi(q), 
  \label{eq:lemma42_0} \\
  & = & \sum_{q = 0}^{\infty} \left( c(\lambda) + \frac{dc(\mu)}{d\mu} \bigg \vert_{\mu = \lambda} \bras{\mu(q) - \lambda} - l(\mu(q)) + G(\mu(q) - \lambda) \right) \pi(q), \nonumber
\end{eqnarray}
where $G(x)$ is a strictly convex function in $x$ as in \cite[Proposition 4.2]{berry}.
We note that $c(\lambda) + \frac{dc(\mu)}{d\mu} \vert_{\mu = \lambda} (\mu(q) - \lambda) = l(\mu(q))$.
Thus we have that $V = \sum_{q = 0}^{\infty} G(\mu(q) - \lambda) \pi(q)$.
From C2, we have that $G(\mu(q) - \lambda) \geq a_{1} (\mu(q) - \lambda)^{2}$ for some constant $a_{1} > 0$.
Thus we have that 
\begin{equation}
  V \geq a_{1} \sum_{q = 0}^{\infty} (\mu(q) - \lambda)^{2} \pi(q).
  \label{chap4:eq:v_variance}
\end{equation}
Consider $q$ such that $\mu(q) \in S$.
Then $\mu(q) - \lambda > \epsilon_{V}$.
Therefore, we have that 
\begin{equation*}
  \sum_{q \in Q_{S}} \epsilon_{V}^{2} \pi(q) \leq \frac{V}{a_{1}},
\end{equation*}
or $Pr\brac{Q \in Q_{s}} = Pr\brac{\mu(Q) \in S} \leq \frac{V}{a_{1} \epsilon_{V}^{2}}$, where $a_{1} > 0$.

Now we consider \INTMC-2.
For \INTMC-2-1 and \INTMC-2-2, let $S_{l} = [0, a_{\lambda} - \epsilon_{V})$ and $S_{r} = (b_{\lambda} + \epsilon_{V}, r_{max}]$.
For ease of exposition, in this proof, for \INTMC-2-3 we set $b_{\lambda} = a_{\lambda} = \lambda$.
For \INTMC-2-3, then $S_{l}$ and $S_{r}$ are as for \INTMC-2-1 and \INTMC-2-2.
From \eqref{eq:lemma42_0} we have that
\begin{eqnarray*}
  V & \geq & \sum_{q \in Q_{S_{l}}} \left( c(\mu(q)) - l(\mu(q))\right) \pi(q) + \sum_{q \in Q_{S_{r}}} \left( c(\mu(q)) - l(\mu(q))\right) \pi(q).
\end{eqnarray*}
Suppose $a_{i} = a_{\lambda}$ and $b_{i} = b_{\lambda}$.
Let $m_{l}$ be the tangent of the angle made by the line through $(a_{i - 1}, c(a_{i - 1}))$ and $(a_{i}, c(a_{i}))$ with $l(\mu)$.
Also let $m_{u}$ be the tangent of the angle made by the line through $(b_{i + 1}, c(b_{i + 1}))$ and $(b_{i}, c(b_{i}))$ with $l(\mu)$.
Then, we note that for $q \in Q_{S_{l}}$, $c(\mu(q)) - l(\mu(q)) \geq m_{l}\brap{\mu(q) - a_{\lambda}}$, while for $q \in Q_{S_{u}}$, $c(\mu(q)) - l(\mu(q)) \geq m_{u}\brap{\mu(q) - b_{\lambda}}$.
Therefore, with $m_{a} = \min(m_{l}, m_{u})$, we have that\footnote{If $a_{i} = 0$, then we note that $m_{a} = m_{u}$.}
\begin{eqnarray*}
  m_{a} \epsilon_{V} \bras{\sum_{q \in Q_{S_{l}}} \pi(q) + \sum_{q \in Q_{S_{u}}} \pi(q)} \leq V.
\end{eqnarray*}
Therefore, we have that $Pr\brac{Q \in Q_{S}} = Pr\brac{\mu(Q) \in S} \leq \frac{V}{m_{a} \epsilon_{V}}$. \qeda

\subsection{Proof of Lemma \ref{prop:p21lb}}
\label{appendix:proof:prop:p21lb}

Consider a particular policy $\gamma$ in the sequence $\gamma_{k}$ with $V_{k} = V$.
Let $\mu^* \stackrel{\Delta} = \lambda - \epsilon_{V}$, where $\epsilon_{V} > 0$ is a function of $V$.
The functional form of $\epsilon_{V}$ will be chosen later.

Let $q_{\mu^*} \Deq \min\brac{q : \mu(q) \geq \mu^{*}}$.
From the non-decreasing property of $\mu(q)$ for $\gamma$, we have that  $Pr\brac{\mu(Q) < \mu^*} = Pr\brac{Q < q_{\mu^*}}$.
Then, from Lemma \ref{lemma:statprobbound} we have that $Pr\brac{\mu(Q) < \mu^*} = \sum_{q = 0}^{q_{\mu^*} - 1} \pi(q) \leq \frac{V}{a_{1} \epsilon_{V}^{2}}$.
We choose $\epsilon_{V}$ as $a_{2}\sqrt{V}$, so that $\alpha \stackrel{\Delta} = \frac{2V}{a_{1} \epsilon_{V}^{2}} = \frac{2}{a_{1} a_{2}^{2}}$. 
We choose $a_{2}$ such that $\alpha < 1$.
In fact, we note that $\alpha$ can be made arbitrarily close to zero by the choice of $a_{2}$.
Therefore, $Pr\brac{\mu(Q) \geq \mu^*} \geq 1 - \frac{\alpha}{2}$, which can be made arbitrarily close to one.

In order to obtain a lower bound on $\overline{Q}(\gamma)$, we intend to find the largest $\overline{q}$ such that $Pr\brac{Q \leq \overline{q}} \leq \frac{1}{2}$.
But we note that $Pr\brac{Q < q_{\mu^*}} \leq \frac{V}{a_{1} \epsilon_{V}^{2}} = \frac{\alpha}{2}$.
Therefore the largest $\overline{q}$ satisfies
\begin{eqnarray*}
  \sum_{q = 0}^{q_{\mu^*} - 1} \pi(q) + \sum_{q = q_{\mu^*}}^{\overline{q}} \pi(q) \leq \frac{1}{2}
\end{eqnarray*}
If $\overline{q}_{1}$ satisfies
\begin{eqnarray*}
  \sum_{q = q_{\mu^*}}^{\overline{q}_{1}} \pi(q) \leq \frac{1}{2} - \frac{\alpha}{2},
\end{eqnarray*}
then $\overline{q}_{1} \leq \overline{q}$, for $\alpha$ sufficiently small.
As $Q(t)$ is a birth-death process, we have that $\pi(q) \lambda = \pi(q + 1) \mu(q + 1)$.
Furthermore, if $q \geq q_{\mu^*}$ we have that $\pi(q - 1) \lambda \geq \pi(q) \mu^*$.
By induction, we obtain that for $q \in \{q_{\mu^*}, \dots\}$
\begin{eqnarray}
  \pi(q) \leq \pi(q_{\mu^*})\left(\frac{\lambda}{\mu^*} \right)^{q - q_{\mu^*}} \leq \pi(q_{\mu^{*}} - 1) \left(\frac{\lambda}{\mu^*}\right)^{q - q_{\mu^*} + 1}, \\
  \text{and for any } q' \geq q_{\mu^*}, \sum_{q = q_{\mu^*}}^{q'} \pi(q) \leq \pi(q_{\mu^*} - 1) \sum_{m = 1}^{q' - q_{\mu^*} + 1} \left(\frac{\lambda}{\mu^*}\right)^{m}.
  \label{chap4:eq:p2-referfromINTLC_1}
\end{eqnarray}
Using the above upper bound on $\sum_{q = q_{\mu^*}}^{q'} \pi(q)$, we obtain a lower bound $\overline{q}_{2}$ to $\overline{q}_{1}$.
If $\overline{q}_{2}$ is the largest integer such that
\begin{eqnarray}
  \pi(q_{\mu^*} - 1) \sum_{m = 1}^{\overline{q}_{2} - q_{\mu^*} + 1} \left(\frac{\lambda}{\mu^*}\right)^{m} \leq \frac{1}{2} - \frac{\alpha}{2},
  \label{chap4:eq:p2-3}
\end{eqnarray}
then $\sum_{q = 0}^{\overline{q}_{2}} \pi(q) \leq \frac{1}{2}$ and $\overline{q}_{1} \geq \overline{q}_{2}$.

Now we obtain an upper bound on $\pi(q_{\mu^*} - 1)$, which is tighter than the upper bound $\frac{\alpha}{2}$ derived before.
From \eqref{chap4:eq:v_variance} we have that
\begin{eqnarray}
  \frac{V}{a_{1}} & \geq & \sum_{q = 0}^{\infty} (\mu(q) - \lambda)^{2} \pi(q) \geq \sum_{q < q_{\mu^*}} (\mu(q) - \lambda)^{2} \pi(q) \nonumber \\
  & = & \sum_{q < q_{\mu^*}} (\mu(q) - \lambda)^{2} \pi(q) + 0 \sum_{q \geq q_{\mu^*}} \pi(q), \nonumber \\
  & \geq & \left( \sum_{q < q_{\mu^*}} (\mu(q) - \lambda) \pi(q) \right)^{2} \text{(using Jensen's inequality as in \cite{berry})}.
\end{eqnarray}
But, as  $\pi(q)\mu(q) = \pi(q - 1)\lambda$, we obtain that
\begin{eqnarray}
  \sum_{q < q_{\mu^*}} (\mu(q) - \lambda) \pi(q) & = & -\lambda \pi(0) + \sum_{1 \leq q \leq q_{\mu^*} - 1} \brap{\lambda\pi(q - 1) - \lambda\pi(q)} = -\lambda \pi(q_{\mu^{*}} - 1), \nonumber \\
  & & \text{ or } \frac{V}{a_{1}} \geq \lambda^{2} \pi(q_{\mu^*} - 1)^{2}.
  \label{chap4:eq:tighterupperbound}
\end{eqnarray}

Now we find a lower bound $\overline{q}_{3}$ on $\overline{q}_{2}$ by using the above upper bound on $\pi(q_{\mu^*} - 1)$ in \eqref{chap4:eq:p2-3}.
Let $\overline{q}_{3}$ be the largest integer such that
\begin{eqnarray*}
  \frac{1}{\lambda}\sqrt{\frac{V}{a_{1}}} \sum_{m = 1}^{\overline{q}_{3} - q_{\mu^*} + 1} \left(\frac{\lambda}{\mu^*}\right)^{m} \leq \frac{1}{2} - \frac{\alpha}{2}.
\end{eqnarray*}
Then $\overline{q}_{3} \leq \overline{q}_{2}$.
We have that $\overline{q}_{3}$ satisfies
\begin{eqnarray*}
  \frac{1}{\lambda}\sqrt{\frac{V}{a_{1}}} \sum_{m = 1}^{\overline{q}_{3} - q_{\mu^*} + 1} \left(\frac{\lambda}{\mu^*}\right)^{m} \leq \frac{1}{2} - \frac{\alpha}{2}, \\
  \frac{\left(\frac{\lambda}{\mu^*}\right)^{\overline{q}_{3} - q_{\mu^*} + 1} - 1}{\lambda - \mu^*} \leq \sqrt{\frac{a_{1}}{V}}\left[ \frac{1 - \alpha}{2} \right], \\
  \left(\frac{\lambda}{\mu^*}\right)^{\overline{q}_{3} - q_{\mu^*} + 1} \leq 1 + \left( \lambda - \mu^* \right)\sqrt{\frac{a_{1}}{V}}\left[ \frac{1 - \alpha}{2} \right] \\
  \overline{q}_{3} - q_{\mu^*} + 1 \leq \log_{\frac{\lambda}{\mu^*}} \left[ 1 + \left( \lambda - \mu^*\right)\sqrt{\frac{a_{1}}{V}}\left[ \frac{1 - \alpha}{2} \right]\right].
\end{eqnarray*}
Since $q_{\mu^*} > 0$, we note that $\overline{q}_{3}$ is at least
\begin{eqnarray*}
  \floor{\log_{\frac{\lambda}{\mu^*}} \left[ 1 + \left( \lambda - \mu^* \right)\sqrt{\frac{a_{1}}{V}}\left[ \frac{1 - \alpha}{2} \right]\right] - 1}.
\end{eqnarray*}
Therefore,
\begin{eqnarray*}
  \overline{q}_{3} & \geq & \log_{\frac{1}{1 - \frac{\epsilon_V}{\lambda}}} \left[ 1 + \epsilon_{V}\sqrt{\frac{a_{1}}{V}}\left[ \frac{1 - \alpha}{2} \right]\right] - 2, \\
  & = & \frac{\log \left[ 1 +  \epsilon_{V}\sqrt{\frac{a_{1}}{V}}\left[ \frac{1 - \alpha}{2} \right]\right]}{-\log(1 - \frac{\epsilon_{V}}{\lambda})} - 2.
\end{eqnarray*}
Since $\epsilon_{V} = a_{2} \sqrt{V}$, we have
\begin{eqnarray*}
  \overline{q}_{3} & \geq & \frac{\log \left[ 1 + a_{2}\sqrt{a_{1}}\left[ \frac{1 - \alpha}{2} \right]\right]}{-\log\brap{1 - \frac{a_{2}\sqrt{V}}{\lambda}}} - 2.
\end{eqnarray*}
Since $\overline{Q}(\gamma) \geq \frac{\overline{q}}{2} \geq \frac{\overline{q}_{1}}{2} \geq \frac{\overline{q}_{2}}{2} \geq \frac{\overline{q}_{3}}{2}$ we have that
\begin{eqnarray*}
  \overline{Q}(\gamma) & \geq & \frac{1}{2} \bras{\frac{\log \left[ 1 + a_{2}\sqrt{a_{1}}\left[ \frac{1 - \alpha}{2} \right]\right]}{-\log\brap{1 - \frac{a_{2}\sqrt{V}}{\lambda}}} - 2 }.
\end{eqnarray*}
As $V \downarrow 0$, we note that $\log\brap{1 - \frac{a_{2}\sqrt{V}}{\lambda}} = \Theta\brap{\sqrt{V}}$.
Hence, for the sequence $\gamma_{k}$ with $\overline{C}(\gamma_{k}) - c(\lambda) = V_{k} \downarrow 0$, we have that $\overline{Q}(\gamma_{k}) = \Omega\nfrac{1}{\sqrt{V_k}}$. 
We consider a sequence $c_{c,k} \downarrow c(\lambda)$.
Let $\gamma_{k,\epsilon}$ be any sequence of $\epsilon$-optimal policies for $c_{c,k}$, for some $\epsilon > 0$.
Then we have that $\overline{Q}(\gamma_{k,\epsilon}) \leq Q^*(c_{c,k}) + \epsilon$.
Therefore, we have that $Q^*(c_{c,k}) = \Omega\nfrac{1}{\sqrt{c_{c,k} - c(\lambda)}}$.
\qeda

\subsection{Proof of Lemma \ref{prop:p221lb}}
\label{appendix:proof:prop:p221lb}
We note that in this case there exists a policy $\gamma^*$, for which the birth death process is irreducible on $\sZ$ and $\mu(q) = b_{1} = b_{\lambda}, \forall q > 0$, with $\overline{C}(\gamma^*) = c(\lambda)$ and $\overline{Q}(\gamma^*) = \frac{\lambda}{b_{\lambda} - \lambda}$.
For $V = 0$, we note that the above policy is optimal.
Consider a particular policy $\gamma$ in the sequence $\gamma_{k}$ with $V_{k} = V$.
Let  $\mu^* = b_{\lambda} + \epsilon_{V}$, where $\epsilon_{V}$ is a function of $V$ to be chosen later.
Then from Lemma \ref{lemma:statprobbound} we have that
\begin{eqnarray}
  \frac{V}{m_{a} \epsilon_{V}} & \geq & \sum_{\mu_{k} \geq \mu^*} \pi_{\mu}(k),
  \label{chap4:eq:p221_1}
\end{eqnarray}
Intuitively, since $\sum_{\mu_{k} \geq \mu^*} \pi_{\mu}(k)$ should approach $0$ as $V \downarrow 0$, we require that the choice of $\epsilon_{V}$ should be such that $\frac{V}{\epsilon_{V}} \downarrow 0$ as $V \downarrow 0$.

Let $q_{\mu^*} \Deq \min\brac{q : \mu(q) \geq \mu^*}$.
For $q < q_{\mu^*}$, $\mu(q) < \mu^*$ and therefore $\pi(q) \lambda < \pi(q + 1)\mu^*$.
Hence, by induction we obtain that
\begin{eqnarray}
  \pi(q_{\mu^*} - m) < \pi(q_{\mu^*}) \left( \frac{\mu^*}{\lambda}\right)^{m}, \text{ for } m \in \{1,\dots,q_{\mu^*}\}.
  \label{chap4:eq:p221_2}
\end{eqnarray}
From \eqref{chap4:eq:p221_1}, we have
\begin{eqnarray*}
  \sum_{q < q_{\mu^*}} \pi(q) = 1 - \sum_{q \geq q_{\mu^*}} \pi(q) \geq 1 - \frac{V}{m_{a} \epsilon_{V}}.
\end{eqnarray*}
Now from \eqref{chap4:eq:p221_2} we have
\begin{eqnarray*}
  \sum_{q < q_{\mu^*}} \pi(q) \leq \sum_{q = 0}^{q_{\mu^*} - 1} \pi(q_{\mu^*}) \left(\frac{\mu^*}{\lambda}\right)^{q_{\mu^*} - q},\text{ hence we have that,} \\
  1 - \frac{V}{m_{a} \epsilon_{V}} \leq \pi(q_{\mu^*}) \sum_{m = 1}^{q_{\mu^*}} \left(\frac{\mu^*}{\lambda}\right)^{m}, \\
  \frac{1}{\pi(q_{\mu^*})} \left( 1 - \frac{V}{m_{a} \epsilon_{V}} \right) \leq \sum_{m = 1}^{q_{\mu^*}} \left(\frac{\mu^*}{\lambda}\right)^{m} = \frac{\mu^*}{\mu^* - \lambda} \left[ \left( \frac{\mu^*}{\lambda} \right)^{q_{\mu^*}} - 1 \right].
\end{eqnarray*}
We note that as $\pi(q_{\mu^*}) \leq \sum_{q \geq q_{\mu^*}} \pi(q) \leq \frac{V}{m_{a} \epsilon_{V}}$, we have that $\frac{1}{\pi(q_{\mu^*})} \geq \frac{m_{a} \epsilon_{V}}{V}$ and therefore
\begin{eqnarray*}
  \frac{m_{a} \epsilon_{V}}{V} \left( 1 - \frac{V}{m_{a} \epsilon_{V}} \right) \leq  \frac{\mu^*}{\mu^* - \lambda} \left[ \left( \frac{\mu^*}{\lambda} \right)^{q_{\mu^*}} - 1 \right], \nonumber \\
  \log_{\frac{\mu^*}{\lambda}}\left[\frac{\mu^* - \lambda}{\mu^*} \frac{m_{a}\epsilon_{V}}{V}\left( 1 - \frac{V}{m_{a} \epsilon_{V}} \right) + 1\right] \leq q_{\mu^*}.
\end{eqnarray*}
By definition, for every $q < q_{\mu^*}$, $\mu(q) < \mu^*$, and for every $q \geq q_{\mu^*}, \mu(q) \leq r_{max}$.
Let us define 
\begin{eqnarray*}
  q_{\mu^*,l} = \left\lceil\log_{\frac{\mu^*}{\lambda}}\left[\frac{\mu^* - \lambda}{\mu^*} \frac{m_{a}\epsilon_{V}}{V}\left( 1 - \frac{V}{m_{a} \epsilon_{V}} \right) + 1\right] \right\rceil,
\end{eqnarray*}
which is the smallest possible value for $q_{\mu^*}$ for any feasible admissible policy $\gamma$.
Consider another policy $\gamma'$ defined as follows :
\begin{eqnarray*}
  \mu(0) & = & 0, \\
  \mu(q) & = & \mu^*, \text{ for } 1 \leq q \leq q_{\mu^*,l}, \\
  \mu(q) & = & r_{max}, \text{ for } q > q_{\mu^*,l}.
\end{eqnarray*}
Then $\overline{Q}(\gamma') \leq \overline{Q}(\gamma)$.
We now obtain a lower bound on $\overline{Q}(\gamma')$.
For $\gamma'$ it can be shown that
\begin{eqnarray*}
  \overline{Q}(\gamma') & \geq & (1 - a)\left[ \frac{a}{(1 - a)^{2}} \left\{1 - (1-a)(q_{\mu^*,l} + 1) a^{q_{\mu^*,l}} - a.a^{q_{\mu^*,l}}\right\} + a^{q_{\mu^*,l}}\left\{ q_{\mu^*,l} \frac{b}{1 - b} + \frac{b}{(1 - b)^{2}} \right\} \right], \\
  & = & \frac{a}{1 - a} + (1 - a)\left[ a^{q_{\mu^*,l}}\left\{ q_{\mu^*,l} \frac{b}{1 - b} + \frac{b}{(1 - b)^{2}}\right\} - \frac{a}{(1 - a)^{2}} \left\{ (1 - a)(q_{\mu^*,l} + 1)a^{q_{\mu^*,l}} + a.a^{q_{\mu^*,l}} \right\}\right], \\
  & \geq & \frac{a}{1 - a} - \frac{a}{1 - a} a^{q_{\mu^*,l}}\left[1 + (1 - a)q_{\mu^*,l}\right],
\end{eqnarray*}
where $a = \frac{\lambda}{\mu^*}$ and $b = \frac{\lambda}{r_{max}}$.
We note that for $V \downarrow 0$, the term $\frac{a}{1 - a} = \frac{\lambda}{b_{\lambda} - \lambda}\brap{1 - \frac{\epsilon_{V}}{b_{\lambda} - \lambda} + o(\epsilon_{V})}$.
If $V \downarrow 0$, since we require that $\frac{V}{\epsilon_{V}} \downarrow 0$, $q_{\mu^*,l} \uparrow \infty$ and therefore the second term in the lower bound for $\overline{Q}(\gamma')$ is $a q_{\mu^*,l} a^{q_{\mu^*,l}}$.
We note that at $V = 0$, since we require that the lower bound is tight, we only consider $\epsilon_{V}$ such that $\epsilon_{V} \downarrow 0$ as $V \downarrow 0$.
Then it can be shown that $\overline{Q}(\gamma) \geq \overline{Q}(\gamma') = \frac{\lambda}{b_{\lambda} - \lambda} - \mathcal{O}\brap{\epsilon_{V} + \frac{V}{\epsilon_{V}}\log\nfrac{\epsilon_{V}}{V}}$, for any sequence $\epsilon_{V} \downarrow 0$ and $\frac{V}{\epsilon_{V}} \downarrow 0$ as $V \downarrow 0$.
By choosing $\epsilon_{V} = {V}^{1 - \delta}$, where $0 < \delta < 1$, we obtain that 
\begin{eqnarray*}
  \overline{Q}(\gamma) = \frac{\lambda}{b_{\lambda} - \lambda} - \mathcal{O}\brap{{V}^{1- \delta}\log\nfrac{1}{V}}.
\end{eqnarray*}
For the sequence $\gamma_{k}$, we therefore obtain that $\overline{Q}(\gamma_{k}) = \frac{\lambda}{b_{\lambda} - \lambda} - \mathcal{O}\brap{{V}^{1- \delta}\log\nfrac{1}{V}}$.
We consider a sequence $c_{c,k} \downarrow c(\lambda)$.
Let $\gamma_{k,\epsilon_{k}}$ be any sequence of $\epsilon_{k}$-optimal policies for $c_{c,k}$, for a sequence $\epsilon_{k}$.
We choose $\epsilon_{k} = c_{c,k} - c(\lambda)$.
Then we have that $\overline{Q}(\gamma_{k,\epsilon}) \leq Q^*(c_{c,k}) + \epsilon_{k}$.
Therefore, we obtain that $Q^*(c_{c,k}) = \frac{\lambda}{b_{\lambda} - \lambda} - \mathcal{O}\brap{(c_{c,k} - c(\lambda))^{1 - \delta}\log\nfrac{1}{c_{c,k} - c(\lambda)}}$, for $0 < \delta < 1$.
\qeda

\subsection{Proof of Lemma \ref{prop:p222lb}}
\label{appendix:proof:prop:p222lb}
Consider a particular policy $\gamma$ in the sequence $\gamma_{k}$ with $V_{k} = V$.
Let $\mu^{*} \stackrel{\Delta} = a_{\lambda} - \epsilon_{V}$.
Define $q_{\mu^*} = \inf\brac{q : \mu(q) \geq \mu^{*}}$.
As $\gamma$ is admissible, we have that $Pr\brac{\mu(Q) < \mu^*} = Pr\brac{Q < q_{\mu^*}}$.
From Lemma \ref{lemma:statprobbound} we have that
\begin{eqnarray*}
  Pr\brac{Q < q_{\mu^*}} & \leq & \frac{V}{m_{a} \epsilon_{V}}, \\
  \text{and } \pi(q_{\mu^*} - 1) & \leq & \frac{V}{m_{a} \epsilon_{V}}.
\end{eqnarray*}
We now choose $\epsilon_{V} = \epsilon$, a positive constant.
To find a lower bound on $\overline{Q}(\gamma)$, in the following, we intend to find the largest $\overline{q}$ such that $\sum_{q = 0}^{\overline{q}} \pi(q) \leq \frac{1}{2}$.
But we note that $Pr\brac{Q < q_{\mu^*}} \leq \frac{V}{m_{a} \epsilon}$ and for any $q < q_{\mu^*}, \pi(q) \leq \frac{V}{m_{a} \epsilon}$.
Therefore, $\pi(q_{\mu^*}) \leq \pi(q_{\mu^*} - 1) \frac{\lambda}{\mu^*} \leq \frac{\lambda V}{m_{a} \epsilon \mu^*}$.
Let $\overline{q}_{1}$ be the largest integer such that
\begin{eqnarray*}
  \sum_{q = q_{\mu^*}}^{\overline{q}} \pi(q) \leq \frac{1}{2} - \frac{V}{m_{a} \epsilon},
\end{eqnarray*}
then $\overline{q}_{1} \leq \overline{q}$.
Proceeding as for problem \INTMC-1, we obtain a lower bound $\overline{q}_{2}$ on $\overline{q}_{1}$ by using an upper bound for $\pi(q)$.
We note that if $q \geq q_{\mu^*}$ we have that $\pi(q - 1) \lambda \geq \pi(q) \mu^*$.
By induction, we obtain that for $q \in \{q_{\mu^*}, \dots\}$
\begin{eqnarray*}
  \pi(q) \leq \pi(q_{\mu^{*}} - 1) \left(\frac{\lambda}{\mu^*}\right)^{q - q_{\mu^*} + 1}, \\
  \text{and for any } q' \geq q_{\mu^*}, \sum_{q = q_{\mu^*}}^{q'} \pi(q) \leq \pi(q_{\mu^*} - 1) \sum_{m = 1}^{q' - q_{\mu^*} + 1} \left(\frac{\lambda}{\mu^*}\right)^{m}.
\end{eqnarray*}
Using the above upper bound on $\sum_{q = q_{\mu^*}}^{q'} \pi(q)$, and $\pi(q_{\mu^*} - 1) \leq \frac{V}{m_{a} \epsilon}$, we obtain the following lower bound $\overline{q}_{2}$ to $\overline{q}_{1}$.

If $\overline{q}_{2}$ is the largest integer such that
\begin{eqnarray*}
  \frac{V}{m_{a} \epsilon} \sum_{m = 1}^{\overline{q}_{2} - q_{\mu^*} + 1} \left(\frac{\lambda}{\mu^*}\right)^{m} \leq \frac{1}{2} - \frac{V}{m_{a} \epsilon},
\end{eqnarray*}
then $\sum_{q = 0}^{\overline{q}_{2}} \pi(q) \leq \frac{1}{2}$ and $\overline{q}_{2} \leq \overline{q}_{1}$.
Hence, $\overline{q}_{2}$ is the largest integer such that
\begin{eqnarray*}
  \left( \frac{\lambda}{a_{\lambda} - \epsilon} \right)^{\overline{q}_{2} - q_{\mu^*} + 1} \leq 1 + \left( \frac{\lambda - a_{\lambda} + \epsilon}{{\lambda}} \right) \frac{\epsilon m_{a}}{V} \left( \frac{1}{2} - \frac{V}{m_{a} \epsilon}\right) \\
  \overline{q}_{2} \leq q_{\mu^*} - 1 + \log_{ \frac{\lambda}{a_{\lambda} - \epsilon}}\left[ 1 + \left( \frac{\lambda - a_{\lambda} + \epsilon}{{\lambda}} \right) \frac{\epsilon m_{a}}{V} \left( \frac{1}{2} - \frac{V}{m_{a} \epsilon}\right)\right].
\end{eqnarray*}
Since $q_{\mu^*} \geq 0$, $\overline{q}_{2}$ is at least
\begin{eqnarray*}
  \floor{\log_{ \frac{\lambda}{a_{\lambda} - \epsilon}}\left[ 1 + \left( \frac{\lambda - a_{\lambda} + \epsilon}{{\lambda}} \right) \frac{\epsilon m_{a}}{V} \left( \frac{1}{2} - \frac{V}{m_{a} \epsilon}\right)\right] - 1}.
\end{eqnarray*}
Therefore,
\begin{eqnarray*}
  \overline{Q}(\gamma) \geq \frac{\overline{q}}{2} \geq \frac{\overline{q}_{1}}{2} \geq \frac{\overline{q}_{2}}{2} \geq \frac{1}{2}\bras{\log_{ \frac{\lambda}{a_{\lambda} - \epsilon}}\left[ 1 + \left( \frac{\lambda - a_{\lambda} + \epsilon}{{\lambda}} \right) \frac{\epsilon m_{a}}{V} \left( \frac{1}{2} - \frac{V}{m_{a} \epsilon}\right)\right] - 2}.
\end{eqnarray*}
Hence, for any sequence $\gamma_{k}$ with $\overline{C}(\gamma_{k}) - c(\lambda) = V_{k} \downarrow 0$, we obtain that $\overline{Q}(\gamma_{k}) = \Omega\left(\log\left(\frac{1}{V_{k}}\right)\right)$.\qeda

\subsection{Proof of Lemma \ref{prop:p223lb}}
\label{appendix:proof:prop:p223lb}

Consider a policy $\gamma$ in the sequence $\gamma_{k}$, with $V_{k} = V$.
Let $\mu^* \stackrel{\Delta} = \lambda - \epsilon_{V} = a_{\lambda} - \epsilon_{V}$.
Define $q_{\mu^*} = \inf\brac{q : \mu(q) \geq \mu^{*}}$.
Since $\gamma$ is admissible, from Lemma \ref{lemma:statprobbound} we have that $Pr\brac{Q < q_{\mu^*}} = Pr\brac{\mu(Q) < \mu^*} \leq \frac{V}{m_{a} \epsilon_V}$.
Let $\epsilon_{V} \stackrel{\Delta} = a_{2} V$, where $a_{2}$ is chosen so that $\alpha \stackrel{\Delta} = \frac{2V}{m_{a} \epsilon_{V}} < 1$.
We note that $a_{2}$ can be chosen such that $\alpha$ is arbitrarily close to zero.
Then, we have
\begin{eqnarray*}
  Pr\brac{Q < q_{\mu^*}} & \leq & \frac{\alpha}{2}, \\
  \text{and } \pi(q_{\mu^*} - 1) & \leq & \frac{\alpha}{2}.
\end{eqnarray*}
To find a lower bound on $\overline{Q}(\gamma)$, in the following, we intend to find the largest $\overline{q}$, such that $\sum_{q = 0}^{\overline{q}} \pi(q) \leq \frac{1}{2}$.
But we note that $Pr\brac{Q < q_{\mu^*}} \leq \frac{\alpha}{2}$ and for any $q < q_{\mu^*}, \pi(q) \leq \frac{\alpha}{2}$.
Therefore, $\pi(q_{\mu^*}) \leq \pi(q_{\mu^*} - 1) \frac{\lambda}{\mu^*} \leq \frac{\lambda \alpha}{2 \mu^*}$.
If $\overline{q}_{1}$ is the largest integer such that
\begin{eqnarray*}
  \sum_{q = q_{\mu^*}}^{\overline{q}} \pi(q) \leq \frac{1}{2} - \frac{V}{m_{a} \epsilon_{V}},
\end{eqnarray*}
then $\overline{q}_{1} \leq \overline{q}$.
Proceeding as for problem \INTMC-1, we obtain a lower bound $\overline{q}_{2}$ on $\overline{q}_{1}$ by using an upper bound for $\pi(q)$.
We note that if $q \geq q_{\mu^*}$ we have that $\pi(q - 1) \lambda \geq \pi(q) \mu^*$.
By induction, we obtain that for $q \in \{q_{\mu^*}, \dots\}$
\begin{eqnarray*}
  \pi(q) \leq \pi(q_{\mu^{*}} - 1) \left(\frac{\lambda}{\mu^*}\right)^{q - q_{\mu^*} + 1}, \\
  \text{and for any } q' \geq q_{\mu^*}, \sum_{q = q_{\mu^*}}^{q'} \pi(q) \leq \pi(q_{\mu^*} - 1) \sum_{m = 1}^{q' - q_{\mu^*} + 1} \left(\frac{\lambda}{\mu^*}\right)^{m}.
\end{eqnarray*}
Using the above upper bound on $\sum_{q = q_{\mu^*}}^{q'} \pi(q)$ we obtain the following lower bound $\overline{q}_{2}$ to $\overline{q}_{1}$.
If $\overline{q}_{2}$ is the largest integer such that
\begin{eqnarray*}
  \pi(q_{\mu^*} - 1) \sum_{m = 1}^{\overline{q}_{2} - q_{\mu^*} + 1} \left( \frac{\lambda}{\mu^*} \right)^{m} \leq \frac{1 - \alpha}{2},
\end{eqnarray*}
then $\overline{q}_{2} \leq \overline{q}$.

We note that 
\begin{eqnarray*}
  V & \geq & \sum_{\mu_{k} < \mu^*} m_{a}(\lambda - \mu_{k}) \pi_{\mu}(k), \\
  & = & \sum_{q < q_{\mu^*}} m_{a}(\lambda - \mu(q))\pi(q).
\end{eqnarray*}
Again, since $\pi(q)\mu(q) = \pi(q - 1)\lambda$, it follows that
\begin{eqnarray*}
  V \geq m_{a} \lambda \pi(q_{\mu^*} - 1).
\end{eqnarray*}
Now if $\overline{q}_{3}$ is the largest integer such that
\begin{eqnarray*}
  \frac{V}{m_{a} \lambda} \sum_{m = 1}^{\overline{q}_{3} - q_{\mu^*} + 1} \left( \frac{\lambda}{\mu^*} \right)^{m} \leq \frac{1 - \alpha}{2},
\end{eqnarray*}
then $\overline{q}_{3} \leq \overline{q}_{2}$.
We have that $\overline{q}_{3}$ satisfies
\begin{eqnarray*}
  \overline{q}_{3} \leq q_{\mu^*} - 1 + \log_{\frac{\lambda}{\mu^*}} \left[ 1 + \frac{\epsilon_{V}}{\lambda} \frac{m_{a} \lambda}{V} \frac{1 - \alpha}{2} \right], \\
  \overline{q}_{3} \leq q_{\mu^*} - 1 + \frac{ \log \left[ 1 + \frac{\epsilon_{V}}{\lambda} \frac{m_{a} \lambda}{V} \frac{1 - \alpha}{2} \right]}{-\log\left(1 - \frac{\epsilon_V}{\lambda}\right)}. 
\end{eqnarray*}
Since $q_{\mu^*} \geq 0$, and $\epsilon_{V} = a_{2}V$, we have that $\overline{q}_{3}$ is at least
\begin{eqnarray*}
  \floor{\frac{ \log \left[ 1 + m_{a} a_{2} \frac{1 - \alpha}{2} \right]}{-\log\left(1 - \frac{a_{2} V}{\lambda}\right)} - 1}.
\end{eqnarray*}
So that 
\begin{eqnarray*}
  \overline{Q}(\gamma) \geq \frac{\overline{q}}{2} \geq \frac{\overline{q}_{1}}{2} \geq \frac{\overline{q}_{2}}{2} \geq \frac{\overline{q}_{3}}{2} \geq \frac{1}{2}\bras{\frac{ \log \left[ 1 + m_{a} a_{2} \frac{1 - \alpha}{2} \right]}{-\log\left(1 - \frac{a_{2} V}{\lambda}\right)} - 2}.
\end{eqnarray*}
Since $\log\left(1 - \frac{a_{2} V}{\lambda}\right) = \Theta\brap{V}$ as $V \downarrow 0$, we have that for any sequence $\gamma_{k}$ with $\overline{C}(\gamma_{k}) - c(\lambda) = V_{k} \downarrow 0$, $\overline{Q}(\gamma_{k}) = \Omega\nfrac{1}{V_{k}}$.\qeda

\subsection{Proof outline for Lemma \ref{prop:p21ub}}
\label{appendix:outline:prop:p21ub}

Each policy $\gamma$ in the sequence of policies $\gamma_{k}$ is defined as follows.
\begin{eqnarray*}
  \mu(0) & = & 0, \\
  \mu(q) & = & \lambda - \epsilon_{U}, \text{ for } q \in \brac{1, \dots, q_{1}}, \\
  \mu(q) & = & \lambda + \epsilon'_{U}, \text{ for } q \in \brac{q_{1} + 1, \dots, 2q_{1}}, \\
  \mu(q) & = & \lambda + K, \text{ for } q \in \brac{2q_{1} + 1, \dots}.
\end{eqnarray*}
Let $\epsilon_{U} = \sqrt{U}$, $\epsilon'_{U} = \frac{\lambda\epsilon_{U}}{\lambda - \epsilon_{U}}$, and $K > 0$ be such that $\lambda + K \leq r_{max}$.
We also let $q_{1} = \floor{\log_{\nfrac{\lambda}{\lambda - \epsilon_{U}}} \brap{1 + \frac{\epsilon_{U}}{U\lambda}}}$.
The sequence of policies $\gamma_{k}$ is obtained by choosing $U$ from a sequence $U_{k} \downarrow 0$.

We obtain $\Qg$ and $\Cg$ for $\gamma$ by evaluating $\pi(q)$.
We have that
\begin{eqnarray*}
  \pi(q) & = 
  \begin{cases}
    \pi(0) \fpow{\lambda}{\lambda - \epsilon_{U}}{q} & \text{ for } q \in \brac{1, \dots, q_{1}}, \\
    \pi(0) \fpow{\lambda}{\lambda - \epsilon_{U}}{q_{1}} \fpow{\lambda}{\lambda + \epsilon'_{U}}{q - q_{1}} & \text{ for } q \in \brac{q_{1} + 1, \dotsm 2q_{1}}, \\
    \pi(0) \fpow{\lambda}{\lambda + K}{q - 2q_{1}} & \text{ for } q \in \brac{2q_{1} + 1, \dots}.
  \end{cases}
\end{eqnarray*}
Since $\sum_{q = 0}^{\infty} \pi(q) = 1$ and $q_{1} \geq \log_{\nfrac{\lambda}{\lambda - \epsilon_{U}}} \brap{1 + \frac{\epsilon_{U}}{U\lambda}} - 1$, it can be shown that $\pi(0) = \mathcal{O}(U)$.
We note that 
\begin{eqnarray*}
  \Cg = \pi(0).0 + \pi_{\mu}(\lambda - \epsilon_{U})c(\lambda - \epsilon_{U}) + \pi_{\mu}(\lambda + \epsilon'_{U}) c(\lambda + \epsilon'_{U}) + \pi_{\mu}(\lambda + K) c(\lambda + K).
\end{eqnarray*}
Then it can be shown that $\Cg \leq c(\lambda) + \mathcal{O}(U)$.
Or if $V = \Cg - c(\lambda)$, then $V = \mathcal{O}(U)$.
Corresponding to the sequence $U_{k}$, we have a sequence $V_{k} = \mathcal{O}(U_{k})$

We note that $\lambda - \mu(q) = K$, for $q > 2q_{1}$.
Then from Proposition \ref{appendix:prop:qlength_ub}, $\Qg = \mathcal{O}\nfrac{2q_{1}}{K}$, or $\Qg = \mathcal{O}\brap{\frac{1}{\sqrt{U}}\log\nfrac{1}{U}}$.
Therefore, we have a sequence of policies $\gamma_{k}$ with $\Qgk = \mathcal{O}\brap{\frac{1}{\sqrt{U_{k}}}\log\nfrac{1}{U_{k}}}$.
Then, we have that $\Qgk = \mathcal{O}\brap{\frac{1}{\sqrt{V_{k}}}\log\nfrac{1}{V_{k}}}$ and $\Cgk - c(\lambda) = V_{k}$.

\subsection{Proof outline for Lemma \ref{prop:p221ub}}
\label{appendix:outline:prop:p221ub}

Each policy $\gamma$ in the sequence of policies $\gamma_{k}$ is defined as follows.
\begin{eqnarray*}
  \mu(0) & = & 0, \\
  \mu(q) & = & b_{\lambda}, \text{ for } q \in \brac{1, \dots, q_{k}}, \\
  \mu(q) & = & r_{max}, \text{ for } q \in \brac{q_{k} + 1, \dots}.
\end{eqnarray*}
The sequence $\gamma_{k}$ is obtained by choosing $q_{k}$ from the sequence $\brac{1, 2, \dots}$.
We evaluate $\Qg$ and $\Cg$ for the policy $\gamma$ by evaluating $\pi(q)$ as in the proof of Lemma \ref{prop:p21ub}.
Then it can be shown that for the sequence $\gamma_{k}$, $\frac{b_{\lambda}}{b_{\lambda} - \lambda} - \Qgk = \Theta\brap{V_{k} \log\nfrac{1}{V_{k}}}$, where $V_{k} = \Cgk - c(\lambda)$.

\subsection{Proof outline for Lemma \ref{prop:p222ub}}
\label{appendix:outline:prop:p222ub}

Each policy $\gamma$ in the sequence of policies $\gamma_{k}$ is defined as follows.
\begin{eqnarray*}
  \mu(0) & = & 0, \\
  \mu(q) & = & a_{\lambda}, \text{ for } q \in \brac{1, \dots, q_{1}}, \\
  \mu(q) & = & b_{\lambda}, \text{ for } q \in \brac{q_{1} + 1, \dots}.
\end{eqnarray*}
We choose $q_{1} = \ceiling{\log_{\nfrac{\lambda}{a_{\lambda}}}\brap{1 + \frac{\lambda - a_{\lambda}}{\lambda}\frac{1}{U}}}$, where $U > 0$.
The sequence $\gamma_{k}$ is obtained by choosing $U$ from a sequence $U_{k} \downarrow 0$.
We again evaluate $\Qg$ and $\Cg$ for the policy $\gamma$ by evaluating $\pi(q)$ as in the proof of Lemma \ref{prop:p21ub}.
Then it can be shown that for the sequence $\gamma_{k}$, $\Qgk = \mathcal{O}\brap{\log\nfrac{1}{V_{k}}}$, where $V_{k} = \Cgk - c(\lambda)$.

\subsection{Proof of Proposition \ref{prop:p222}}
\label{appendix:proof:prop:p222}

Let $V_{k} = \Cgk - c(\lambda)$ for the sequence of policies $\gamma_{k}$ as in the proposition.
Let $\gamma_{\epsilon,k}$ be any sequence of $\epsilon$-optimal policies for $c_{c,k} = \Cgk$ for some $\epsilon > 0$.
Then we have that $\overline{Q}\brap{\gamma_{\epsilon,k}} \leq Q^*(c_{c,k}) + \epsilon$.
Or we have that $Q^*(c_{c,k}) = \overline{Q}\brap{\gamma_{\epsilon,k}} - \epsilon$.
From Lemma \ref{prop:p222lb} we have that $\overline{Q}\brap{\gamma_{\epsilon,k}} = \Omega\brap{\log\nfrac{1}{V_{k}}}$.
Therefore, $Q^*(c_{c,k}) =  \Omega\brap{\log\nfrac{1}{c_{c,k} - c(\lambda)}}$.
Combining this with the asymptotic upper bound from Lemma \ref{prop:p222ub} we have that $Q^*(c_{c,k}) =  \Theta\brap{\log\nfrac{1}{c_{c,k} - c(\lambda)}}$.\qeda

\subsection{Proof outline for Lemma \ref{lemma:p223ub}}
\label{appendix:outline:p223ub}

Each policy $\gamma$ in the sequence of policies $\gamma_{k}$ is defined as follows.
\begin{eqnarray*}
  \mu(0) & = & 0, \\
  \mu(q) & = & \lambda, \text{ for } q \in \brac{1, \dots, q_{1}}, \\
  \mu(q) & = & \lambda + K, \text{ for } q \in \brac{q_{1} + 1, \dots}.
\end{eqnarray*}
We note that $\lambda = a_{i}$ for $i > 1$.
Then $K$ is chosen such that $\lambda + K \leq a_{i + 1}$.
We choose $q_{1} = \ceiling{\frac{1}{U}}$, where $U > 0$.
The sequence $\gamma_{k}$ is obtained by choosing $U$ from a sequence $U_{k} \downarrow 0$.
We again evaluate $\Qg$ and $\Cg$ for the policy $\gamma$ by evaluating $\pi(q)$ as in the proof of Lemma \ref{prop:p21ub}.
Then it can be shown that for the sequence $\gamma_{k}$, $\Qgk = \mathcal{O}\nfrac{1}{V_{k}}$, where $V_{k} = \Cgk - c(\lambda)$.

\subsection{Proof of Lemma \ref{lemma:intlc_pi0ub}}
\label{appendix:proof:lemma:intlc_pi0ub}
Consider a particular policy $\gamma$ in the sequence with $V_{k} = V$.
We note that $\Ug \leq u(\Exp \lambda(Q))$.
Since $\gamma$ is admissible, we have that $\Ug \leq u(\Exp \mu(Q))$.
We then have that $u^{-1}(\Ug) \leq \Exp \mu(Q) = (1 - \pi(0)) \mu$, since $u^{-1}(.)$ exists if $u(\lambda)$ is concave and increasing in $\lambda$.
Therefore, we have that
\begin{eqnarray*}
  \pi(0) \leq 1 - \frac{u^{-1}(\Ug)}{\mu} = 1 - \frac{u^{-1}(u(\mu) - V)}{\mu}.
\end{eqnarray*}
We note that $u^{-1}(x) \geq l^{-1}(x)$, $x \in \sR$, where $l^{-1}(.)$ is the inverse function of $l(\lambda)$.
Then we have 
\begin{eqnarray*}
  \pi(0) \leq 1 - \frac{l^{-1}(u(\mu) - V)}{\mu} = 1 - \frac{l^{-1}(u(\mu)) - mV}{\mu} = \frac{mV}{\mu},
\end{eqnarray*}
since $u(\mu) = l(\mu)$ and where $m$ is the slope of $l^{-1}$.
Therefore, for the sequence $\gamma_{k}$, $\pi(0) = \mathcal{O}(V_{k})$.
For \INTLC, as $u_{c} \uparrow u(\mu)$, for any sequence $\gamma_{k}$ of feasible policies, $u(\mu) - \Ugk \downarrow 0$ and hence $\pi(0) \downarrow 0$. \qeda

\subsection{Proof of Lemma \ref{lemma:statprobbound:lc}}
\label{appendix:proof:lemma:statprobbound:lc}

The proof is very similar to that of Lemma \ref{lemma:statprobbound}.
We consider \INTLC-1 first.
As in the proof of Lemma \ref{prop:p21lb}, we have that 
\begin{eqnarray*}
  V & = & \sum_{q = 0}^{\infty} \pi(q) \bras{\brap{\mu - \lambda(q)}\frac{du(\lambda)}{d\lambda}\vert_{\lambda = \mu} + G(\lambda(q) - \mu)},
\end{eqnarray*}
where $G(x)$ is as in the proof of Lemma \ref{lemma:statprobbound}.
From U2, we have that there exists a positive $a_{1}$ such that 
\begin{eqnarray*}
  V & \geq & \sum_{q = 0}^{\infty} \pi(q) \bras{\brap{\mu - \lambda(q)}\frac{du(\lambda)}{d\lambda}\vert_{\lambda = \mu}  + a_{1}(\lambda(q) - \mu)^{2}}.
\end{eqnarray*}
Since $\sum_{q = 0}^{\infty} \pi(q)\lambda(q) \leq \mu$, we have that
\begin{eqnarray*}
  V & \geq & \sum_{q = 0}^{\infty} \pi(q) a_{1}(\lambda(q) - \mu)^{2}.
\end{eqnarray*}
The rest of the proof for \INTLC-1 is similar to that of Lemma \ref{lemma:statprobbound}.
For \INTLC-2, we define $S_{l} = [0, a_{\mu} - \epsilon_{V})$ and $S_{u} = (b_{\mu} + \epsilon_{V}, r_{a,max}]$ and proceed as in the case of \INTLMC-2 to obtain the bounds as stated in the lemma.
\qeda

\subsection{Proof of Lemma \ref{lemma:intlc-1:lb}}
\label{appendix:proof:lemma:intlc-1:lb}
The proof is similar to that of Lemma \ref{prop:p21lb}, but with some minor differences.
We again consider a particular policy in the sequence with $V_{k} = V$.
Let $\lambda^* \Deq  \mu + \epsilon_{V}$, where $\epsilon_{V} > 0$ is a function of $V$ to be chosen later.
Let $q_{\lambda^*} \Deq \min\brac{q : \lambda(q) \leq \lambda^*}$.
We note that unlike $q_{\mu^*}$ in Lemma \ref{prop:p21lb}, $q_{\lambda^*}$ could be $0$.
We proceed as in the proof of Lemma \ref{prop:p21lb} by choosing $\epsilon_{V} = a_{2} \sqrt{V}$.
Then from Lemma \ref{lemma:statprobbound:lc} $Pr\brac{Q < q_{\lambda^*}} \leq \frac{1}{a_{1}a_{2}}$ if $q_{\lambda^*} > 0$.
As before, we choose $a_{2}$ such that $Pr\brac{Q < q_{\lambda^*}} \leq \frac{\alpha}{2}$, where $\alpha$ can be made arbitrarily close to zero.
If $q_{\lambda^*} = 0$, then $Pr\brac{Q < q_{\lambda^*}} = 0 \leq \frac{1}{a_{1}a_{2}}$.

As in the proof of Lemma \ref{prop:p21lb} we find the largest $\overline{q}$ such that $Pr\brac{Q \leq \overline{q}} \leq \frac{1}{2}$.
We note that if $q \geq q_{\lambda^*}$, then $\pi(q - 1)\lambda^* \geq \pi(q) \mu$.
Then by induction we obtain that for any $q \geq q_{\lambda^*}$,
\begin{eqnarray}
  \sum_{q = q_{\lambda^*}}^{q} \pi(q) \leq \pi(q_{\lambda^*}) \sum_{m = 0}^{q - q_{\lambda^*}} \fpow{\lambda^*}{\mu}{m}.
  \label{chap4:eq:intlc_indeq}
\end{eqnarray}
We note that this is similar to \eqref{chap4:eq:p2-referfromINTLC_1}, except that we express the above upper bound in terms of $\pi(q_{\lambda^*})$ rather than $\pi(q_{\mu^*} - 1)$ in \eqref{chap4:eq:p2-referfromINTLC_1}, since $q_{\lambda^*}$ could be zero.

If $q_{\lambda^*} = 0$, then from Lemma \ref{lemma:intlc_pi0ub} we have that $\pi(q_{\lambda^*}) = \pi(0) = \mathcal{O}(V)$.
If $q_{\lambda^*} > 0$, we obtain an upper bound on $\pi(q_{\lambda^*})$, as in the proof of Lemma \ref{prop:p21lb}.
We have that 
\begin{eqnarray*}
  \frac{V}{a_{1}} & \geq & \sum_{q < q_{\lambda^*}} \pi(q) (\lambda(q) - \mu)^{2} \geq \brap{\sum_{q < q_{\lambda^* - 1}}(\lambda(q) - \mu)\pi(q)}^{2}.
\end{eqnarray*}
Since for $q > 0$, since $\pi(q)\lambda(q) = \mu \pi(q + 1)$, we proceed as in the proof of Lemma \ref{prop:p21lb} to obtain that
\begin{eqnarray*}
  \frac{V}{a_{1}} & \geq & \brap{\mu \pi(q_{\lambda^*}) - \mu \pi(0)}^{2}, \\
  & = & \mu^{2} \pi(q_{\lambda^*})^{2} + \mu^{2} \pi(0)^{2} - 2\mu^{2} \pi(q_{\lambda^*}) \pi(0).
\end{eqnarray*}
Since $\pi(0) \geq 0$ and $\pi(0) = \mathcal{O}(V)$ from Lemma \ref{lemma:intlc_pi0ub}, we have that
\begin{eqnarray*}
  \frac{V}{a_{1}} + 2\mu^{2} \pi(q_{\lambda^*}) \pi(0) & \geq & \mu^{2} \pi(q_{\lambda^*})^{2}, \\
  \frac{V}{a_{1}} + 2\mu^{2} \mathcal{O}(V) & \geq & \mu^{2} \pi(q_{\lambda^*})^{2}, \\
  \text{or } \pi(q_{\lambda^*}) & = & \mathcal{O}(\sqrt{V}).
\end{eqnarray*}
We note that for both $q_{\lambda^*} = 0$ or $q_{\lambda^*} > 0$, we have that $\pi(q_{\lambda^*}) = \mathcal{O}(\sqrt{V})$.

We now proceed as in the proof of Lemma \ref{prop:p21lb}, by using \eqref{chap4:eq:intlc_indeq}, to find the largest integer $\overline{q}$ such that 
\begin{eqnarray*}
  \pi(q_{\lambda^*}) \sum_{m = 0}^{q - q_{\lambda^*}} \fpow{\lambda^*}{\mu}{m} \leq \frac{1}{2} - \frac{\alpha}{2}.
\end{eqnarray*}
The rest of the proof is similar to that of Lemma \ref{prop:p21lb}, and we obtain that $\overline{Q}(\gamma_{k}) = \Omega\nfrac{1}{\sqrt{V_{k}}}$.
Then given a sequence of $u_{c,k} \uparrow u(\mu)$, we have that there exists a sequence of $\epsilon$-optimal $\gamma_{k}$ such that $\Qgk \leq Q^*(u_{c,k}) + \epsilon$, for some $\epsilon > 0$.
Therefore, $Q^*(u_{c,k}) = \Omega\nfrac{1}{\sqrt{u(\mu) - u_{c,k}}}$, since $u_{c,k} \leq \Ugk$.\qeda

\subsection{Proof of Lemma \ref{lemma:intlc-2-1:lb}}
\label{appendix:proof:lemma:intlc-2-1:lb}
We first consider the case $i < P$, where $i$ is such that $b_{i} = b_{\mu}$.
The proof follows that of Lemma \ref{prop:p222lb}.
We define $\lambda^* = b_{\mu} + \epsilon$, where $\epsilon > 0$.
Let $q_{\lambda^*} \Deq \min\brac{q : \lambda(q) \leq \lambda^*}$.
We note that $q_{\lambda^*}$ could be $0$, unlike $q_{\mu^*}$ in Lemma \ref{prop:p222lb}.
If $q_{\lambda^*} = 0$, then we have that $\pi(q_{\lambda^*}) = \pi(0) = \mathcal{O}(V)$.
If $q_{\lambda^*} > 0$, then from Lemma \ref{lemma:statprobbound:lc} we have that $Pr\brac{Q < q_{\lambda^{*}}} \leq \frac{V}{m_{a}\epsilon}$ and $\pi(q_{\lambda^*} - 1) \leq \frac{V}{m_{a} \epsilon}$.
Since $\pi(q_{\lambda^*} - 1)\lambda(q_{\lambda^*} - 1) = \pi(q_{\lambda^*})\mu$ we have that $\pi(q_{\lambda^*}) \leq \pi(q_{\lambda^*} - 1) \frac{r_{a,max}}{\mu}$.
We note that therefore $\pi(q_{\lambda^*}) = \mathcal{O}(V)$ for both $q_{\lambda^*} = 0$ and $q_{\lambda^*} > 0$.

Now proceeding as in the proof of Lemma \ref{prop:p222lb} we have that for any $q \geq q_{\lambda^*}$ (we express the bound in terms of $\pi(q_{\lambda^*})$)
\begin{eqnarray}
  \sum_{q = q_{\lambda^*}}^{q} \pi(q) \leq \pi(q_{\lambda^*}) \sum_{m = 0}^{q - q_{\lambda^*}} \fpow{\lambda^*}{\mu}{m}.
  \label{chap4:eq:intlc_21_1}
\end{eqnarray}
We note that independently of whether $q_{\lambda^*}$ is $0$ or not, if we find the largest $\overline{q}$ such that
\begin{eqnarray*}
  \sum_{q = q_{\lambda^*}}^{\overline{q}} \pi(q) \leq \frac{1}{2} - \frac{V}{a_{1} \epsilon},
\end{eqnarray*}
then $\overline{Q}(\gamma) \geq \frac{\overline{q}}{2}$.
We now proceed as in the proof of Lemma \ref{prop:p222lb}, using the upper bound in \eqref{chap4:eq:intlc_21_1} and $\pi(q_{\lambda^*}) = \mathcal{O}(V)$ to obtain that $\overline{Q}(\gamma_{k}) = \Omega\brap{\log\nfrac{1}{V_{k}}}$.
Now given a sequence of $u_{c,k} \uparrow u(\mu)$, we have that there exists a sequence of $\epsilon$-optimal $\gamma_{k}$ such that $\Qgk \leq Q^*(u_{c,k}) + \epsilon$, for some $\epsilon > 0$.
Therefore, $Q^*(u_{c,k}) = \Omega\brap{\log\nfrac{1}{u(\mu) - u_{c,k}}}$, since $u_{c,k} \leq \Ugk$.

Now we consider the case $i = P$.
We consider \INTLC-2-1 for a larger $\mathcal{X}_{\lambda}$ defined as follows.
We extend $\mathcal{X}_{\lambda}$ to $\overline{\mathcal{X}}_{\lambda} = \mathcal{X}_{\lambda} \cup (b_{P}, b_{P} + \delta]$, for some $\delta > 0$.
We also extend the definition of $u(.)$ to $\overline{\mathcal{X}}_{\lambda}$, by choosing a piecewise linear function on $(b_{P}, b_{P} + \delta]$ which preserves the strictly increasing concave property of $u(.)$.
We denote $Q^*(u_{c})$ when $\lambda(q) \in \overline{\mathcal{X}}_{\lambda}$, by $Q^*_{e}(u_{c})$.
Then we note that $Q^*_{e}(u_{c}) \leq Q^*(u_{c})$.
The asymptotic lower bound for $Q^*_{e}(u_{c})$ follows from the above derivation, which then also holds for $Q^*(u_{c})$.
\qeda

\subsection{Proof of Lemma \ref{lemma:intlc-2-2:lb}}
\label{appendix:proof:lemma:intlc-2-2:lb}
The proof is similar to that of Lemma \ref{prop:p223lb}.
We choose $\lambda^* = \mu + \epsilon_{V} = a_{\mu} + \epsilon_{V}$.
Let $q_{\lambda^*} \Deq \min\brac{q : \lambda(q) \leq \lambda^*}$.
We note that $q_{\lambda^*}$ could be $0$, unlike $q_{\mu^*}$ in Lemma \ref{prop:p223lb}.
If $q_{\lambda^*}$ is $0$, then we note that $\pi(q_{\lambda^*}) = \pi(0) = \mathcal{O}(V)$ from Lemma \ref{lemma:intlc_pi0ub}.
If $q_{\lambda^*} > 0$, then from Lemma \ref{lemma:statprobbound:lc} we have that $Pr\brac{Q < q_{\lambda^*}} \leq \frac{V}{m_{a} \epsilon_{V}}$ and $\pi(q_{\lambda^*} - 1) \leq \frac{V}{m_{a} \epsilon_{V}}$.
We also note that since $\pi(q_{\lambda^*} - 1)\lambda(q_{\lambda^*} - 1) = \pi(q_{\lambda^*})\mu$ we have that $\pi(q_{\lambda^*}) \leq \pi(q_{\lambda^*} - 1) \frac{r_{a,max}}{\mu}$.
Therefore $\pi(q_{\lambda^*}) \leq \frac{V}{m_{a}\epsilon_{V}} \frac{r_{a,max}}{\mu}$.
We choose $\epsilon_{V} = a_{2} V$, so that $\frac{r_{a,max}}{m_{a} \mu a_{2}} \leq \frac{\alpha}{2}$.
We note that $a_{2}$ can be chosen such that $\alpha$ is arbitrarily small.

Then, as in the proof of Lemma \ref{prop:p223lb}, if $\overline{q}$ is the largest integer such that 
\[ \sum_{q = q_{\lambda^*}}^{\overline{q}} \pi(q) \leq \frac{1}{2} - \frac{\alpha}{2},\]
then $\overline{Q}(\gamma) \geq \frac{\overline{q}}{2}$, independently of whether $q_{\lambda^*} = 0$ or not.

We note that for any $q \geq q_{\lambda^*}$ we have that
\begin{eqnarray}
  \sum_{q = q_{\lambda^*}}^{q} \pi(q) \leq \pi(q_{\lambda^*}) \sum_{m = 0}^{q - q_{\lambda^*}} \fpow{\lambda^*}{\mu}{m}.
  \label{chap4:eq:intlc_2231}
\end{eqnarray}

We also note that if $q_{\lambda^*} > 0$, then we have that
\begin{eqnarray*}
  V & \geq & \sum_{\lambda_{k} > \lambda^*} m_{a} \brap{\lambda_{k} - \mu}\pi_{\lambda}(k), \\
  & = & m_{a} \sum_{q < q_{\lambda^*}} \brap{\lambda(q) - \mu}\pi(q), \\
  & = & m_{a} \mu\pi(q_{\lambda^*}) - \mu\pi(0).
\end{eqnarray*}
Then since $\pi(0) = \mathcal{O}(V)$ we have that $\pi(q_{\lambda^*}) = \mathcal{O}(V)$.
Thus independently of whether $q_{\lambda^*} = 0$ or not, we have that $\pi(q_{\lambda^*}) = \mathcal{O}(V)$.

Now proceeding as in the proof of Lemma \ref{prop:p223lb}, using the above upper bound on $\pi(q_{\lambda^*})$ in \eqref{chap4:eq:intlc_2231} we have that $\overline{Q}(\gamma_{k}) = \Omega\nfrac{1}{V_{k}}$.
Now given a sequence of $u_{c,k} \uparrow u(\mu)$, we have that there exists a sequence of $\epsilon$-optimal $\gamma_{k}$ such that $\Qgk \leq Q^*(u_{c,k}) + \epsilon$, for some $\epsilon > 0$.
Therefore, $Q^*(u_{c,k}) = \Omega\nfrac{1}{{u(\mu) - u_{c,k}}}$, since $u_{c,k} \leq \Ugk$.\qeda

\subsection{Proof of Lemma \ref{prop:p3lb}}
\label{appendix:proof:prop:p3lb}
We consider a particular policy $\gamma$ in the sequence $\gamma_{k}$ with $V_{k} = V$.
Since $\gamma$ is admissible, we have that $\Expp \mu(Q) = \Expp \lambda(Q)$.
From the concavity of $u(\lambda)$, we have that $\Expp \lambda(Q) \geq \newl$.
Let $\mu^* = \newl - \epsilon_{V}$, where $\epv$ is a function of $V$ to be chosen later.
Define $\zms \Deq \min\brac{q : \mu(q) \geq \mu^*}$.
We note that $\forall q < \zms$, $\mu(q) < \mu^*$.
As $\mu(q)$ is non-decreasing, we have that
\begin{eqnarray*}
  Pr\brac{ Q < \zms} = Pr\brac{\mu(Q) < \mu^*}.
\end{eqnarray*}
Let the countable set of service rates be denoted by $\brac{\mu_{0} = 0, \mu_{1}, \dots}$, where $\mu_{i} < \mu_{i + 1}$ and $\mu_{i} \in [0,r_{max}]$.
Let $l(\mu)$ be the tangent line at $(\newl, c(\newl))$ to the curve $c(\mu)$.
Then $V = \sum_{q = 0}^{\infty} \bras{c(\mz) - l(\mz)}\pi(q)$.
Proceeding as in Lemma \ref{lemma:statprobbound} we have that $\exists a_{1} > 0$ such that
\begin{eqnarray*}
  Pr\brac{Q \leq \zms - 1} \leq \frac{V}{a_{1}\epv^{2}}, \\
  \text{and } \pi(\zms - 1) \leq \frac{V}{a_{1}\epv^{2}}.
\end{eqnarray*}
Now, since $Q(t)$ is a birth death process $\forall q$, we have that $\pi(q) \lz = \pi(q + 1)\mu(q + 1)$.
For any $q \geq \zms$,
\begin{eqnarray}
  \pi(q + 1) = \frac{\pi(q)\lz}{\mu(q + 1)} \leq \frac{\pi(q) r_{a,max}}{\mu^*}, \nonumber \\
  \pi(q) \leq \pi(\zms - 1) \fpow{r_{a,max}}{\mu^*}{q - \zms + 1}.
  \label{chap4:eq:p31}
\end{eqnarray}
Let $\overline{q}$ be the largest integer such that $\sum_{q = 0}^{\overline{q}} \pi(q) \leq \frac{1}{2}$.
We find a lower bound on $\overline{q}$ as in the proof of Lemma \ref{prop:p222lb}.
We note that $Pr\brac{Q \geq \zms} \geq 1 - \frac{V}{a_{1}\epv^{2}}$.
Let $\epv = \epsilon$, where $0 < \epsilon < u^{-1}(u_{c})$.
For $V$ small, let $\overline{q}_{1}$ be the largest integer such that
\begin{eqnarray*}
  \sum_{q = \zms}^{\overline{q}_{1}} \pi(q) \leq \frac{1}{2} - \frac{V}{a_{1}\epv^{2}}.
\end{eqnarray*}
Then $\overline{q}_{1} \leq \overline{q}$.
We find a lower bound on $\overline{q}_{1}$ by using the upper bound on $\pi(q)$ from \eqref{chap4:eq:p31}.
Let $\overline{q}_{2}$ be the largest integer such that
\begin{eqnarray*}
  \pi(\zms - 1) \sum_{q = 1}^{\overline{q}_{2} - \zms + 1} \fpow{r_{a,max}}{\mu^*}{q} \leq \frac{1}{2} - \frac{V}{a_{1}\epv^{2}}.
\end{eqnarray*}
Then $\overline{q}_{2} \leq \overline{q}_{1}$.
After substituting for $\mu^*$, we have that any $\overline{q}_{2}$ satisfying the above inequality is such that
\begin{eqnarray*}
  \overline{q}_{2} - \zms + 1 \leq \log_{\nfrac{r_{a,max}}{\newl - \epv}} \brap{1 + \frac{r_{a,max} - \newl + \epv}{r_{a,max}} \frac{1}{\pi(\zms - 1)}\brap{\frac{1}{2} - \frac{V}{a_{1}\epv^{2}}}}.
\end{eqnarray*}
Hence we obtain that $\overline{q}_{2}$ is at least
\begin{eqnarray*}
  \log_{\nfrac{r_{a,max}}{\newl - \epv}} \brap{1 + \frac{r_{a,max} - \newl + \epv}{r_{a,max}} \frac{1}{\pi(\zms - 1)}\brap{\frac{1}{2} - \frac{V}{a_{1}\epv^{2}}}} - 2.
\end{eqnarray*}
We note that $\frac{1}{\pi(\zms - 1)} \geq \frac{a_{1}\epsilon^{2}}{V}$ and is the dominant term in the regime where $V \downarrow 0$.
Since $\overline{q} \geq \overline{q}_{1} \geq \overline{q}_{2}$ and $\overline{Q}(\gamma) \geq \frac{\overline{q}}{2}$, we have that for any sequence of $\gamma_{k}$ with $\overline{C}(\gamma_{k}) - \newl = V_{k} \downarrow 0$, $\overline{Q}(\gamma_{k}) = \Omega\brap{\log\nfrac{1}{V_{k}}}$.\qeda

\subsection{Proof of Lemma \ref{prop:p3ub}}
\label{appendix:proof:prop:p3ub}
\renewcommand{\epv}{\epsilon_{U}}

Consider a policy $\gamma$ of the following form:
\begin{eqnarray*}
  &	\mu(0)  & =  0, \\
  & 	\mu(q)  & = \muo = \newl - \epv, \text{ for } q \in \brac{1,\dots,q_{1}}, \\
  &	\mu(q)  & = \mut = \newl + \epv, \text{ for } q \in \brac{q_{1} + 1,\dots}; \\
  \text{and } & \lambda(q) & = \lo, \text{ for } q \in \brac{0, \dots, q_{1} - 1}, \\
  & \lambda(q) & = \newl, \text{ for } q \in \brac{q_{1}, \dots, q_{1} + K}, \\
  & \lambda(q) & = \lt, \text{ for } q \in \brac{q_{1} + K + 1, \dots}.
\end{eqnarray*}
Let $\epv = U$, $\lo > \newl > \lt, \lo > \muo, \lt < \mut$, and $q_{1} = \left \lceil \log_{\nfrac{\lambdao}{\mo}} \brap{1 + \frac{\lambdao - \mo}{\lambdao}\frac{{1}}{U}} \right \rceil$,
We will specify $K$, $\lambda_{1}$, and $\lambda_{2}$ later.
Let $\frac{dc(\newl)}{d\mu} \stackrel{\Delta} = \frac{dc(\mu)}{d\mu}\vert_{\mu = \newl}$. 
We now obtain $\overline{C}(\gamma)$.
\begin{eqnarray*}
  \overline{C}(\gamma) & = & \pi(0).0 + \pi_{\mu}(\mo) c(\mo) + \pi_{\mu}(\mt) c(\mt), \\
  & = & \pi_{\mu}(\mo) \brap{c(\newl) + (-\epv)\frac{dc(\newl)}{d\mu} + \mathcal{O}(\epv^{2})} + \\ 
  & & \pi_{\mu}(\mt) \brap{c(\newl) + (\epv)\frac{dc(\newl)}{d\mu} + \mathcal{O}(\epv^{2})}, \nonumber \\
  & \leq & c(\newl) + \mathcal{O}(U^{2}) + (-\epv \pi_{\mu}(\mo) + \epv \pi_{\mu}(\mt))\frac{dc(\newl)}{d\mu}, \\
  & \leq & c(\newl) + \epv\frac{dc(\newl)}{d\mu} + \mathcal{O}(U^{2}), \\
  & \leq & c(\newl) + \mathcal{O}(U),
\end{eqnarray*}
where $\frac{dc(\newl)}{d\mu} \Deq \frac{dc(\mu)}{d\mu}\vert_{\mu = \newl}$.
Let $V = \overline{C}(\gamma) - c(\newl)$, then we have that $V = \mathcal{O}(U)$.
For $\gamma$, the average utility is
\begin{eqnarray*}
  \overline{U}(\gamma) = u(\lo)\brap{ \sum_{q = 0}^{q_{1} - 1} \pi(q) } + u_{c} \sum_{q = q_{1}}^{q_{1} + K} \pi(q) + u(\lt) \sum_{q = q_{1} + K + 1}^{\infty} \pi(q).
\end{eqnarray*}
Let $\lo = \newl + \epsilon$ and $\lt = \newl - \epsilon$, where $\epsilon$ is a small positive constant.
Then for strictly concave and linear $u(.)$ we have that
\begin{eqnarray*}
  \overline{U}(\gamma) & \geq & u_{c} + \bras{\epsilon \sum_{q = 0}^{q_{1} - 1} \pi(q) - \epsilon \sum_{q = q_{1} + K + 1}^{\infty} \pi(q)}D(u(\newl)) + O(u(\newl)),
\end{eqnarray*}
where $D(u(\newl))$ and $O(u(\newl))$ are defined as follows.
If $u(.)$ is a strictly concave function, then it is differentiable at $\newl$ and $D(u(\newl)) \Deq \frac{du(\lambda)}{d\lambda}\vert_{\lambda = \newl}$ and $O(u(\newl)) = \mathcal{O}(\epsilon^{2})$, with the above inequality being an equality.
If $u(.)$ is linear then $D(u(\newl)) \Deq \frac{du(\lambda)}{d\lambda}\vert_{\lambda = \newl}$ and $O(u(\newl)) = 0$, with the above inequality being an equality.

In the following we show that $\sum_{q = q_{1} + K + 1}^{\infty} \pi(q) \leq \sum_{q = 0}^{q_{1} - 1} \pi(q)$, in which case we have that $\overline{U}(\gamma) \geq u_{c}$ for sufficiently small $\epsilon$ (which is fixed and independent of $V$).
We have that
\begin{eqnarray*}
  \pi(q) = \pi(0) \fpow{\lambdao}{\muo}{q}, & \text{ for } q \in \brac{1,\dots,q_{1}}, \\
  \pi(q) = \pi(0) \fpow{\lo}{\mo}{q_{1}} \fpow{\newl}{\mt}{q - q_{1}}, & \text{ for } q \in \brac{q_{1} + 1, \dots, q_{1} + K}, \\
  \pi(q) = \pi(0) \fpow{\lambdao}{\muo}{q_{1}}\fpow{\newl}{\mt}{K}\nfrac{\newl}{\mt}\fpow{\lambdat}{\mut}{q - q_{1} - K - 1}, & \text{ for } q \in \brac{q_{1} + K + 1,\dots}.
\end{eqnarray*}
Therefore, 
\begin{eqnarray*}
  \sum_{q = 0}^{q_{1} - 1} \pi(q)  & = & \pi(0) + \pi(0)\sum_{q = 1}^{q_{1} - 1} \fpow{\lo}{\mo}{q}, \\
  & = & \pi(0) + \pi(0)\nfrac{\lo}{\lo - \mo}\brap{\fpow{\lo}{\mo}{q_{1} - 1} - 1}.
\end{eqnarray*}
And,
\begin{eqnarray*}
  \sum_{q_{1} + K + 1}^{\infty}\pi(q) & = & \pi(0)\fpow{\lambdao}{\muo}{q_{1}}\fpow{\newl}{\mt}{K}\nfrac{\newl}{\mt}\sum_{q = 0}^{\infty}\fpow{\lt}{\mt}{q}, \\
  & = & \pi(0)\fpow{\lambdao}{\muo}{q_{1}}\fpow{\newl}{\mt}{K}\nfrac{\newl}{\mt - \lt}.
\end{eqnarray*}
If 
\begin{eqnarray}
  \fpow{\lambdao}{\muo}{q_{1}}\fpow{\newl}{\mt}{K}\nfrac{\newl}{\mt - \lt} \leq \nfrac{\lo}{\lo - \mo}\brap{\fpow{\lo}{\mo}{q_{1} - 1} - 1} + 1
  \label{chap4:eq:p3s}
\end{eqnarray}
then $\sum_{q = q_{1} + K + 1}^{\infty} \pi(q) \leq \sum_{q = 0}^{q_{1} - 1} \pi(q)$.
We note that \eqref{chap4:eq:p3s} can be simplified to the question
\begin{eqnarray*}
  \frac{\mo}{\lo - \mo} \stackrel{?} \leq \fpow{\lo}{\mo}{q_1}\brap{\frac{\mo}{\lo - \mo} - \fpow{\newl}{\mt}{K}\nfrac{\newl}{\mt - \lt}}, \\
  \frac{\mo}{\lo - \mo} \stackrel{?} \leq \fpow{\lo}{\mo}{q_1}\brap{\frac{\newl - \epv}{\epsilon + \epv} - \fpow{\newl}{\newl + \epv}{K}\nfrac{\newl}{\epsilon + \epv}}, \\
  \frac{\mo}{\lo - \mo} \stackrel{?} \leq \fpow{\lo}{\mo}{q_1}\nfrac{\newl}{\epsilon + \epv}\brap{1 - \frac{\epv}{\newl} - \brap{1 + \frac{\epv}{\newl}}^{-K}}.
\end{eqnarray*}
We use the lower bound on $q_{1}$, obtained by removing the ceiling, to arrive at the following question :
\begin{eqnarray*}
  \frac{\newl - \epv}{\epsilon + \epv} \stackrel{?} \leq \brap{1 + \frac{\epsilon + \epv}{\newl - \epv}\frac{1}{U}} \nfrac{\newl}{\epsilon + \epv}\brap{1 - \frac{\epv}{\newl} - \brap{1 + \frac{\epv}{\newl}}^{-K}}
\end{eqnarray*}
For sufficiently small $U$, with $\epv = U$, we have that $\brap{1 + \frac{\epv}{\newl}}^{-K} \leq 1 - \frac{K\epv}{2\newl}$.
So, instead of the above question we can ask the stronger question
\begin{eqnarray*}
  \frac{\newl - \epv}{\epsilon + \epv} \stackrel{?} \leq \brap{1 + \frac{\epsilon + \epv}{\newl - \epv}\frac{1}{V}} \nfrac{\newl}{\epsilon + \epv}\brap{\frac{\epv}{\newl}\brap{\frac{K}{2} - 1}}.
\end{eqnarray*}
We choose $K > 2\brap{1 + \frac{(\newl)^{2}}{\epsilon}}$.
Then we can ask the even stronger questions
\begin{eqnarray*}
  \newl - \epv \stackrel{?} \leq \brap{1 + \frac{\epsilon + \epv}{\newl - \epv}\frac{1}{V}}\brap{\epv\frac{(\newl)^{2}}{\epsilon}}, \\
  \newl - \epv \stackrel{?} \leq \frac{\epsilon + \epv}{\epsilon}\frac{(\newl)^{2}}{\newl - \epv}
\end{eqnarray*}
which indeed hold.
Hence for sufficiently small $V$ and $\epsilon$, $\overline{U}(\gamma) \geq u_{c}$.
From Proposition \ref{appendix:prop:qlength_ub} we have
\begin{eqnarray*}
  \overline{Q}(\gamma) = \frac{(q_{1} + 1)(\epv + \epsilon + r_{a.max})}{\epv + \epsilon} + \frac{r_{max}}{2(\epv + \epsilon)}, \\
  \overline{Q}(\gamma) \leq \frac{(q_{1} + 1)(\epsilon + r_{a,max})}{\epsilon} + \frac{r_{max}}{2\epsilon}.
\end{eqnarray*}
As $q_{1} = \mathcal{O}\brap{\log\nfrac{1}{U}}$, we obtain that $\overline{Q}(\gamma) = \mathcal{O}\brap{\log\nfrac{1}{U}}$.
We note that the policy $\gamma$ is admissible.
The sequence of policies is obtained by choosing $U_{k} = \frac{1}{k}$.
We note that then we have a corresponding sequence $V_{k} = \mathcal{O}(U_{k})$.
Thus, $\overline{Q}(\gamma_{k}) = \mathcal{O}\brap{\log\nfrac{1}{V_{k}}}$, and we have that there exists a sequence of admissible policies $\gamma_{k}$ with a corresponding sequence $V_{k} \downarrow 0$ such that
\begin{eqnarray*}
  \overline{Q}(\gamma_{k}) & = & \mathcal{O}\brap{\log\nfrac{1}{V_{k}}},\\
  \overline{C}(\gamma_{k}) - c(\newl) & = & V_{k}, \\
  \overline{U}(\gamma_{k}) & \geq & u_{c}.
\end{eqnarray*}\qeda

\subsection{Proof of equivalence of SP2(b) and SP2(a) in Section \ref{chap4:sec:sim_probs_INTMC}}
\label{appendix:proof:sp2b_sp2a}

We now show that the asymptotic regime for SP2(b) is equivalent to that for SP2(a), i.e., $c_{c,k} - c\brap{u^{-1}(u_{c,k})} \downarrow 0$.
We note that for any $(u_{c,k})$ and $(c_{c,k})$, for which the problem \eqref{chap4:eq:intlmc_problem_statement} is feasible, and also such that $u(c^{-1}(c_{c,k})) - u_{c,k} \downarrow 0$, we have that $\forall \epsilon > 0$, $\exists K_{\epsilon}$ such that, $\forall k > K_{\epsilon}$, $u(c^{-1}(c_{c,k})) - \epsilon \leq u_{c,k} \leq u(c^{-1}(c_{c,k}))$ (since $u_{c,k} \leq u(c^{-1}(c_{c,k}))$ if the problem \eqref{chap4:eq:intlmc_problem_statement} is feasible).
Then we have that $u^{-1}\brap{u(c^{-1}(c_{c,k})) - \epsilon} \leq u^{-1}(u_{c,k}) \leq c^{-1}(c_{c,k})$.
For every $c_{c,k}$, we define $l_{1,k}(\lambda)$ to be (i) the tangent to $u(\lambda)$ at $(c^{-1}(c_{c,k}), u\brap{c^{-1}(c_{c,k})} )$, if $u(\lambda)$ is strictly convex, (ii) the line passing through $(a_{\mu}, u(a_{\mu}))$ and $(b_{\mu}, u(b_{\mu}))$, if $u(\lambda)$ is piecewise linear and $c^{-1}(c_{c,k})$ lies on a linear segment, and (iii) any line through $(a_{\mu}, u(a_{\mu}))$ with slope $m$, such that $\frac{du(\lambda)}{d\lambda}^{-}\vert_{\lambda = \mu} < m < \frac{du(\lambda)}{d\lambda}^{+}\vert_{\lambda = \mu}$, if $u(\lambda)$ is piecewise linear and $c^{-1}(c_{c,k})$ is a corner point of $u(\lambda)$.
We note that $l_{1,k}(c^{-1}(c_{c,k})) = u(c^{-1}(c_{c,k}))$ in all three cases.
Then 
\[u^{-1}\brap{u(c^{-1}(c_{c,k})) - \epsilon} \geq l_{1,k}^{-1}\brap{u(c^{-1}(c_{c,k})) - \epsilon} = l_{1,k}^{-1}\brap{u(c^{-1}(c_{c,k}))} - m_{1,k}\epsilon,\]
where $m_{1,k}$ is the slope of $l_{1,k}^{-1}$.
Since $l_{1,k}^{-1}\brap{u(c^{-1}(c_{c,k}))} = c^{-1}(c_{c,k})$, we have that
\begin{eqnarray*}
  c^{-1}(c_{c,k}) - m_{1,k}\epsilon \leq u^{-1}(u_{c,k}) \leq c^{-1}(c_{c,k}), \\
  c\brap{c^{-1}(c_{c,k}) - m_{1,k}\epsilon} \leq c(u^{-1}(u_{c,k})) \leq c_{c,k}.
\end{eqnarray*}
Let $l_{2,k}(\mu)$ be the tangent to $c(\mu)$ at $(c^{-1}(c_{c,k}), c_{c,k})$.
Then we have that
\begin{eqnarray*}
  l_{2,k}\brap{c^{-1}(c_{c,k}) - m_{1,k}\epsilon} \leq c(u^{-1}(u_{c,k})) \leq c_{c,k}, \\
  c_{c,k} - m_{2,k} m_{1,k} \epsilon \leq c(u^{-1}(u_{c,k})) \leq c_{c,k},
\end{eqnarray*}
where $m_{2,k}$ is the slope of $l_{2,k}$.
We note that $\exists M_{1}, M_{2} \in \sR$ such that $m_{1,k} \leq M_{1}$ and $m_{2,k} \leq M_{2}$ for every $k$, since both $u(\lambda)$ and $c(\mu)$ are defined on bounded domains.
Since the above inequality holds for every $\epsilon > 0$ and $k > K_{\epsilon}$, we have that that $c_{c,k} - c(u^{-1}(u_{c,k})) \downarrow 0$.

\subsection{Proof of Lemma \ref{lemma:intlmc-altregime}}
\label{appendix:proof:lemma:intlmc-altregime}
The proof follows that of Lemma \ref{prop:p3lb} closely.
Hence, we only state the differences here.
We define $\mu^* = u^{-1}(u_{c,k}) - \epsilon_{V}$ and $q_{\mu^*} = \inf\brac{q : \mu(q) \geq \mu^*}$.
We note that unlike in the proof of Lemma \ref{prop:p3lb}, we define a different tangent line $l_{k}(\mu)$ for every $u_{c,k}$.
Let $l_{k}(\mu)$ be the tangent line to $c(\mu)$ at $(u^{-1}(u_{c,k}), c(u^{-1}(u_{c,k})))$.
From U2, we have a positive $a_{1,k}$ such that
\begin{eqnarray*}
  V_{k} \geq a_{1,k} \sum_{q = 0}^{\zms - 1} \bras{\mz - u^{-1}(u_{c,k})}^{2} \pi(q).
\end{eqnarray*}
We note that unlike the proof of Lemma \ref{prop:p3lb}, here $a_{1,k}$ depends on the sequence $u_{c,k}$.
Let $a \Deq \inf_{k} \brac{a_{1,k}}$.
Since $u(.)$ is $m$-strongly convex, we have that $a \geq m > 0$.
Then we have that $Pr\brac{Q \leq \zms - 1} \leq \frac{V}{a\epv^{2}}$ and $\pi(\zms - 1) \leq \frac{V}{a\epv^{2}}$.
Then, we proceed as in the proof of Lemma \ref{prop:p3lb} to obtain that $\overline{Q}(\gamma_{k}) = \Omega\brap{\log\nfrac{1}{V_{k}}}$.\qeda

\subsection{Proof of Proposition \ref{prop:duallowerbound}}
\label{appendix:proof:prop:duallowerbound}

\newcommand{\tbk}{\tilde{\beta}_{k}}
\newcommand{\tb}{\tilde{\beta}}

Let $\gamma_{k}$ be any sequence of policies such that $\Cgk - c(\lambda) \downarrow 0$ and $\Qgk = \mathcal{O}\brap{\log\nfrac{1}{\Cgk - c(\lambda)}}$.
Let $\tilde{\beta}_{k} = 1/(\Cgk - c(\lambda))$.
Consider the problem
\begin{eqnarray}
  \min_{\gamma} \brac{\Qg + \tilde{\beta}_{k} \brap{\Cg - c(\lambda)}},
  \label{eq:duallowerbound0}
\end{eqnarray}
for a particular value of $k$.
Let $\gamma^*_{\tbk}$ be an admissible optimal policy for \eqref{eq:duallowerbound0} (we know from \cite{ata} that such an admissible optimal policy exists).
We show that $\overline{C}(\gamma^*_{\tbk}) - c(\lambda) = \mathcal{O}\brap{1/\tbk^{\delta}}$, where $0 < \delta < 1$.
We proceed by contradiction.
Suppose $\overline{C}(\gamma^*_{\tbk}) - c(\lambda) = \omega\brap{1/\tbk^{\delta}}$.
Therefore, the optimal value of \eqref{eq:duallowerbound0}, $\overline{Q}\brap{\gamma^*_{\tbk}} + \tbk\brap{\overline{C}\brap{\gamma^*_{\tbk}} - c(\lambda)} = \omega\brap{\tbk^{1 - \delta}}$.
However, we note that the sequence $\gamma_{k}$ is such that $\Qgk + \tbk\brap{\Cgk - c(\lambda)} = \mathcal{O}\brap{\log\tbk}$, which contradicts the optimality of $\gamma^*_{\tbk}$.
Therefore, $\overline{C}(\gamma^*_{\tbk}) - c(\lambda) = \mathcal{O}\brap{1/\tbk^{\delta}}$.
Now consider a sequence $c_{c,k}$ for \eqref{introduction:eq:mm1_genTradeoff}.
Suppose $c_{c,k} - c(\lambda) = \Theta\brap{\Cgk - c(\lambda)}$.
Then, the Lagrange dual of \eqref{introduction:eq:mm1_genTradeoff} can be bounded below as follows.
\begin{eqnarray*}
  \max_{\beta_{1} \geq 0} \bras{  \min_{\gamma} \Qg + \tilde{\beta}_{k} \brap{\Cg - c_{c,k}}} \geq \overline{Q}\brap{\gamma^*_{\tbk}} + \tbk\brap{\overline{C}\brap{\gamma^*_{\tbk}} - c(\lambda)} - \tbk\brap{c_{c,k}  - c(\lambda)}.
\end{eqnarray*}
We have that $\tbk\brap{\overline{C}\brap{\gamma^*_{\tbk}} - c(\lambda)} \geq 0$. 
Since $c_{c,k} - c(\lambda) = \mathcal{O}\brap{\Cgk - c(\lambda)}$, we have that $\tbk\brap{c_{c,k}  - c(\lambda)} = \mathcal{O}(1)$.
Furthermore, since $\overline{C}(\gamma^*_{\tbk}) - c(\lambda) = \mathcal{O}\brap{1/\tbk^{\delta}}$, we have that $\overline{Q}(\gamma^*_{\tbk}) = \Omega\brap{\log \tbk}$.
Since $c_{c,k} - c(\lambda)$ is also $\Omega\brap{\Cgk - c(\lambda)}$, we have that 
\begin{eqnarray*}
  \overline{Q}\brap{\gamma^*_{\tbk}} + \tbk\brap{\overline{C}\brap{\gamma^*_{\tbk}} - c(\lambda)} - \tbk\brap{c_{c,k}  - c(\lambda)} = \Omega\brap{\log\nfrac{1}{c_{c,k} - c(\lambda)}}.
\end{eqnarray*}
\qeda

\section{Additional result}
The following upper bound on the average queue length is obtained via a Lyapunov drift argument.
\begin{proposition}
Assume that the admissible policy $\gamma$ is such that the birth death process is irreducible on $\sZ$ and there exists a $q_{\epsilon}$ such that $\mu(q_{\epsilon}) - \lambda(q_{\epsilon}) = \epsilon > 0$. Then
\begin{eqnarray*}
\Qg \leq \frac{q_{\epsilon}\brap{\epsilon + r_{a,max}}}{\epsilon} + \frac{r_{max} + r_{a,max}}{2\epsilon}.
\end{eqnarray*}
\label{appendix:prop:qlength_ub}
\end{proposition}
The complete proof is presented in \cite[Proposition 2.A.1]{vineeth_thesis}.

\end{appendices}

\end{document}